\documentclass[3p,sort&compress]{elsarticle}
\usepackage[utf8]{inputenc}
\usepackage{amsmath}
\usepackage{amsfonts}
\usepackage{amssymb}
\usepackage{graphicx}
\usepackage{physics}
\usepackage{booktabs}
\usepackage{siunitx}
\DeclareSIUnit\fps{fps}
\DeclareSIUnit\pixel{pixel}
\usepackage{xcolor}
\journal{arXiv}
\usepackage[hidelinks]{hyperref}
\DeclareGraphicsExtensions{.png,PNG,.pdf,.PDF}
\usepackage{pdfpages}

\title{Interplay between powder catchment efficiency and layer height in self-stabilized laser metal deposition}
\begin{document} 
	
	\begin{frontmatter}
		
		\author[polimi,inrim]{Simone Donadello}
		
		\author[polimi]{Valentina Furlan\corref{corresponding}}
		\cortext[corresponding]{Corresponding author}
		%\ead{valentina.furlan@polimi.it}
		
		\author[polimi]{Ali G\"{o}khan Demir}
		\author[polimi]{\mbox{Barbara Previtali}}
		
		\address[polimi]{Department of Mechanical Engineering, Politecnico di Milano, \mbox{Via La Masa 1, 20156 Milan, Italy}}
		\address[inrim]{Istituto Nazionale di Ricerca Metrologica, INRIM, \mbox{Strada delle Cacce 91, 10135 Turin, Italy}}
		
		\begin{abstract}
			In laser metal deposition (LMD) the height of the deposited track can vary within and between layers, causing significant deviations during the process evolution. Previous works have shown that in certain conditions a self-stabilizing mechanism occurs, maintaining a regular height growth and a constant standoff distance between the part and the deposition nozzle. Here we analyze the link between the powder catchment efficiency and the deposition height stability. To this purpose, a monitoring system was developed to study the deposition in different process conditions, using inline measurements of the specimen weight in combination with the layer height information obtained with coaxial optical triangulation. An analytical model was used to estimate the deposition efficiency in real-time from the height monitoring and the process parameters, which was verified by the direct mass measurements. The results show that the track height stabilization is associated to a reduction of the powder catchment efficiency, which is governed by the melt pool relative position with respect to the powder cone and the laser beam. For a given set of parameters, the standoff distance can be estimated to achieve the highest powder catchment efficiency and a regular height through the build direction.
		\end{abstract}
		
		
	\end{frontmatter}
	\let\thefootnote\relax\footnotetext{Published version: \href{https://doi.org/10.1016/j.optlaseng.2021.106817}{doi.org/10.1016/j.optlaseng.2021.106817}}

	\section{Introduction}
	Laser metal deposition (LMD) is an additive manufacturing (AM) process belonging to the directed energy deposition (DED) family \cite{dass_state_2019}. In the conventional LMD implementation, a laser beam creates a melt pool that captures the coaxial powder stream which is blown to the deposition region, generating the deposition tracks. Complex and large parts with variable slicing axis can be made by the consecutive deposition of adjacent tracks and layers \cite{debroy_additive_2018, flynn_hybrid_2016}. LMD is also capable to produce parts with multi-material deposition \cite{shah_experimental_2014}, build features on existing components, and repair damaged parts \cite{saboori_application_2019}. Despite these several advantages, the LMD process is still not widely applied on an industrial scale. Some of the key issues limiting its industrial application are the limited geometrical stability and the requirement of post-processing \cite{bruzzo_surface_2021}. In LMD the deposited track dimensions vary as a function of the process parameters, but can also depend on the deposition trajectory and part geometry. While, for instance, in laser powder bed fusion the layer height relies on the powder bed lowering, in LMD the deposition track height can vary with the same set of parameters and material due to the thermal history of the process.
	
	In industrial practice, the LMD process parameters are set for achieving mainly pore and crack free deposits \cite{shamsaei_overview_2015}. However, when these parameters are used for generating complex geometries, they can often fail to respect the required geometrical tolerances \cite{eisenbarth_geometry-based_2020}. Such kind of problem may rise from heat accumulation during the process in acute corners or the decrease of the scanned section area, where the deposit height can increase compared to the nominal one. Inversely, the process may derive to a colder stage with long deposition tracks: in this case, the deposit height can be lower than expected and the part may fail to grow. One aspect which is directly correlated to the laser parameters is the cooling rate. For example, a change in the parameter set can increase the temperature gradient between the deposition site and the substrate. Since the main cooling mechanism in this technology is conduction, a higher thermal gradient involves a higher cooling rate. This can translate in a different temperature profile along the deposited layers over the substrate \cite{thompson_overview_2015}. Indeed temperature can influence the deposition growth. At the initial layers the deposition is carried out close to the substrate, which behaves as a heat sink with a lower initial temperature \cite{costa_rapid_2005}. As the deposition proceeds over the layers, the conduction behavior changes due to the heat buildup, which also may affect the material microstructure, as well as the porosity formation and the layer thickness uniformity.
	
	Thermal effects can be mitigated by choosing the optimal set of process parameters from preliminary experimental campaigns or numerical simulations \cite{pirch_laser-aided_2019}. While this approach is effective in resolving bigger issues such as part failure, offline optimization can be time-consuming and does not allow the system to autonomously operate: the actual process fluctuations related to the powder-laser interaction remain uncontrolled, leading to unrepeatable results. Therefore industrial LMD systems can operate with closed-loop controllers, commonly relying on coaxial pyrometers \cite{song_control_2012}, digital cameras \cite{fathi_clad_2007}, or composite systems \cite{tang_layer--layer_2011}. These devices aim to maintain a required process temperature and stable deposition conditions. However they require careful calibration methods, which can also differ from bulky to thin structures. Other inline monitoring devices such as optical triangulation \cite{heralic_height_2012} or low-coherence interferometry \cite{kogel-hollacher_oct_2020} have been also proposed for a direct measurement of the deposit height, which can be compensated by the machine axis movement. Despite several achievements to control these changes in the deposited track height, more attention is required for developing a fundamental understanding of the process physics.
	
	The material deposition stability is strictly related to the capacity of the process to maintain the rate of powder being fused. For instance, the LDM process is characterized by higher resolution but lower deposition efficiency if compared to its counterpart that uses wire feedstock, the laser metal wire deposition \cite{syed_comparative_2005}. In particular, powder catchment efficiency is a key parameter that defines the amount of material being deposited in comparison with the amount being released to the deposition area. This relies on the interplay between powder jet, laser beam, and laser process parameters, which determine the deposition track geometry \cite{paul_laser_2013}. Powder catchment efficiency has been one of the earliest research questions of LMD \cite{lin_simple_1999}. The main importance has been given to the material usage, as the main aim has been to identify the conditions for an environmentally and economically viable process \cite{cacace_using_2020}. For more expensive alloys, such as Ti- and Ni-alloys, the material cost can become a significant fraction of the overall operation costs in LMD based production. The optimization of the laser-powder interaction enables new high-speed deposition technologies \cite{li_extreme_2019}. Nonetheless, the powder catchment efficiency can be a key element for understanding the deposition stability. Recently a method for the powder flow measurement by a precision scale has been demonstrated and correlated to the actual deposition efficiency measured from single tracks \cite{eisenbarth_spatial_2019}. On the other hand, to the authors' knowledge no previous work attempted to measure the powder catchment efficiency in real time on multiple-track and multiple-layer depositions.
	
	The layer height in LMD is known to vary starting from the first deposition stages, since the initial layers are characterized by a certain kind of process instability \cite{donadello_monitoring_2019}. This behavior is determined by layer growth rate variability given by the mismatch between deposition height and part design, which can move on a stable value during the deposition progress \cite{pinkerton_significance_2004,zhu_influence_2012}. Such self-regulation mechanism can be related to a change in the powder catchment efficiency during the deposition: this can eventually converge to a stationary regime, although generating transitory geometric inaccuracies \cite{haley_working_2019}. Accordingly different works faced up to the influence of powder distribution and defocusing on height variability and deposition efficiency while building multi-layer parts \cite{tan_process_2018,lin_process_2020}. However an extensive study and modeling regarding the interplay between deposition growth and powder catchment efficiency is still required for a full comprehension of deposition stabilization and optimization. This way, the phenomena behind the geometrical deviations, especially in the deposition height, can be revealed and compensated throughout the process evolution. 
	
	This work presents an innovative approach to study the powder catchment efficiency in LMD for multi-track multi-layer parts in real-time. The role of deposition height in the mechanism of passive process stabilization was investigated. An analytical model was developed to link the powder catchment efficiency to the deposition height growth and to the main process parameters. An inline experimental setup was developed to measure the mass increase of the deposit during the process. The deposition height was simultaneously measured via coaxial optical triangulation. Different process conditions were considered for the deposition of AISI 316L cubic samples using a LMD setup based on a three-jet nozzle and a multimode fiber laser. The measurements obtained from the multiple instruments were used in a complementary way to identify the factors that influence the powder catchment efficiency. The results clearly indicated that self-regulation is associated to the powder catchment efficiency reduction. In the future, the proposed model might be used to identify the optimal parameters in terms of both process stability and efficiency, and it can be functional for an active control of the deposition accuracy by means of dimensional optical monitoring.

	{	\begin{center}
			
			\footnotesize
			\begin{tabular}{|ll|}
				\hline
				\multicolumn{2}{|l|}{Nomenclature}                                           \\ \hline
				$\eta_{m}$         & powder catchment efficiency                              \\
				$\eta_{m}^*$       & effective powder catchment efficiency                    \\
				$\eta_{path}$      & deposition path coefficient for $\eta_{m}$                        \\
				$\eta_{h}$         & geometrical powder catchment efficiency                       \\
				$\eta_{th}$        & theoretical powder catchment efficiency                       \\
				$\eta_{en}$        & energetic coefficient for $\eta_{th}$                   \\
				$\eta_{int}$       & powder-laser interaction coefficient for $\eta_{th}$    \\
				$\dot{m}_{dep}$    & deposition mass rate                             \\
				$\dot{m}_{tot}$    & total delivered powder mass flow rate                             \\
				$\Delta t_{dep}$   & total deposition time interval                                        \\
				$\Delta t_{proc}$    & processing interval during $\Delta t_{dep}$                \\
				$A_{cs}$           & single track cross-section area              \\
				$\bar h$                & average layer height                                 \\
				$h_{th}$           & theoretical layer height                     \\
				$S$                & standoff distance                                       \\
				$S_0$              & reference standoff distance                             \\
				$S_{th}$           & theoretical standoff distance                           \\
				$r_x$, $r_z$       & robot path transverse and height increments           \\
				$\Delta h$         & height mismatch between $\bar h$ and $r_z$                        \\
				$v$                & transverse scanning speed                               \\
				$P$                & processing laser power                                  \\
				$P_0$              & critical power for powder melting                                   \\
				$\rho$             & solid material density                                  \\
				$\mathcal{C}_p$              & solid material specific heat capacity                           \\
				$\mathcal{L}$                & material fusion latent heat                             \\
				$\mathcal{A}$      & material optical absorptance                                    \\
				$T_{m}$            & material melting point                                  \\
				$\Delta T$         & temperature increase during deposition          \\
				$A_p$              & powder spot area                        \\
				$A_{int}$          & laser-powder interaction area        \\
				$d_{p0}$, $x_{p0}$           & diameter and radial distance of powder orifices          \\
				$x_{p1}$, $z_{p1}$ & extrapolated coordinates of powder cone apex               \\
				$\alpha_p$, $\beta_p$         & half-divergence and inclination of powder jets                       \\
				$x_p$, $x_l$       & coordinates of powder and laser spot borders                \\
				$c_g$              & geometrical correction factor to $x_l$                  \\
				$w_{l0}$, $\theta_l$           & laser beam waist radius and half-divergence                                   \\
				$z_{l0}$           & laser focus axial coordinate                            \\ \hline
			\end{tabular} 
		\end{center}
	}

	\section{Model}
	In the following sections the deposition efficiency will be discussed from different point of views. Section \ref{sec:model1} provides a general definition of the powder catchment efficiency in terms of the deposited mass. In section \ref{sec:model2} a model for the layer height provides an alternative definition of powder catchment efficiency from geometrical considerations. Finally, section \ref{sec:model3} presents a semi-empirical model for the interpretation of the deposition process stability, where the powder catchment efficiency is expressed as a function of the process parameters and of the standoff distance. Such theoretical background links the self-stabilization mechanisms in the deposition height growth to the powder catchment efficiency variability, determined by the variable interaction between the powder and laser beams.

	\subsection{Definition of powder catchment efficiency}
	\label{sec:model1}
	The powder catchment efficiency $\eta_{m}$ is defined as the ratio between the deposition mass rate $\dot{m}_{dep}$, i.e. the increase of the deposit mass in a given time interval, and the total mass flow rate $\dot{m}_{tot}$ of the metallic powder which gets delivered during the deposition:
	\begin{equation}
	\eta_{m} = \frac{\dot{m}_{dep}}{\dot{m}_{tot}}\,.
	\label{eq:eta-general}
	\end{equation}
	The total powder flow $\dot{m}_{tot}$ includes losses caused by inefficiency in the process of powder catchment, thus $\eta_{m}<1$.
	
	A direct estimation of $\eta_{m}$ can be obtained by measuring the mass $\Delta m_{dep}$ deposited during a finite time interval $\Delta t_{dep}$, e.g. the interval for the deposition of a layer. The average efficiency over the corresponding interval is therefore
	\begin{equation}
	\bar\eta_{m} = \frac{\Delta \bar m_{dep}}{\Delta t_{dep} \, \dot{m}_{tot}}.
	\label{eq:eta-w-1}
	\end{equation}
	
	The deposition procedure for non-trivial geometries can include intervals where the deposition is suspended. For example, in robotized systems, the laser emission can be disabled during the position settlement of the deposition head between different tracks or layers. Conversely the powder feed is typically kept constant in order to maintain a stationary flow. Of course the powder which gets delivered when the processing laser is not emitting cannot be deposited and gets lost. Indeed the average powder catchment efficiency defined in Eq. \eqref{eq:eta-w-1} includes an implicit coefficient $\eta_{path}$ that is not related to the process itself, but which is set by the chosen deposition path strategy. This can be expressed as the ratio between the actual processing time $\Delta t_{proc}$ over the total deposition time interval $\Delta t_{dep}$:
	\begin{equation}
	\bar\eta_{path} = \frac{\Delta t_{proc}}{\Delta t_{dep}}.
	\label{eq:eta-path}
	\end{equation}
	Therefore, in real-world measurements, the average efficiency $\bar\eta_{m}$ must be normalized to $\bar\eta_{path}$ to eliminate the dependence on the specific deposition strategy, hence introducing the effective powder catchment efficiency $\bar\eta_{m}^*$ related only to the process physics and defined as
	\begin{equation}
	\bar \eta_{m}^* = \frac{\bar\eta_{m}}{\bar\eta_{path}}\,.
	\label{eq:eta-w-eff}
	\end{equation}

	\subsection{Calculation of efficiency from deposition layer height}
	\label{sec:model2}
	The powder catchment efficiency can be defined also from geometrical considerations. Assuming the absence of porosity, the deposition mass rate is related to the cross-section area $A_{cs}$ of a single deposition track as
	\begin{equation}
	\dot{m}_{dep} = A_{cs} v \rho\,,
	\label{eq:m-dep-acs}
	\end{equation}
	with $v$ being the transverse scanning speed and $\rho$ the bulk material density.
	
\begin{figure}[htb]
		\begin{center}
			\includegraphics[width=0.5\linewidth]{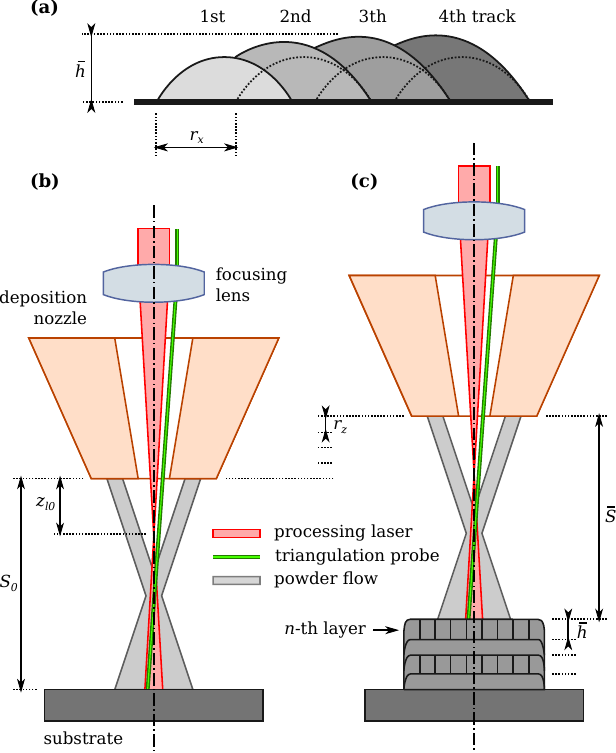}
		\end{center}
		\caption{(a) Cross-section representation of partially overlapping deposition tracks. (b-c) Schematic representation related to the multi-layer deposition height, in the initial condition and after the deposition of some layers.}
		\label{fig:nozzle-track-scheme}
	\end{figure}

	We consider the deposition of a simple prismatic geometry composed of multiple tracks and multiple layers, referred to as bulk deposition from hereon. Each layer is composed of several adjacent tracks separated by a path transverse increment $r_x$, which is assumed constant in the current discussion. In common practice each track partially overlaps with the previous one: this translates in an actual height growth excess, and inclined tracks are typically observed in multi-track LMD \cite{huang_new_2019, ocelik_geometry_2014,sun_-process_2020}. The general experimental evidence is characterized by an initial transient in the track shape, which converges to a stationary regime within few adjacent tracks. From purely geometrical considerations, discussed in the Supplementary Material, the average contribution given by each track to the deposition layer is equivalent to a rectangular cross-section, whose dimensions are approximated by the path transverse increment $r_x$ and by the average layer height $\bar h$ as represented in Figure \ref{fig:nozzle-track-scheme}(a):
	\begin{equation}
	\bar A_{cs} \simeq r_x \bar h\,.
	\label{eq:area-cross-2}
	\end{equation}
	It follows that the powder catchment efficiency can be calculated in terms of geometrical parameters by inserting  Eq. \eqref{eq:m-dep-acs} into the general definition of Eq. \eqref{eq:eta-general}, and taking the average cross-section area of Eq. \eqref{eq:area-cross-2} over a multi-track layer. Therefore, if the deposition parameters $r_x$, $\dot{m}_{tot}$ and $v$ are fixed, and if the deposition process is stationary within each layer, $\bar \eta_{h}$ depends only on the average layer height $\bar h$:
	\begin{equation}
	\bar \eta_{h} \simeq \bar h \frac{r_x v \rho}{\dot{m}_{tot}}\,.
	\label{eq:eta-h-2}
	\end{equation}
	
	It is important to highlight that the geometrical definition of $\bar \eta_{h}$ is equivalent to the general one of $\bar \eta^*_{m}$ given in Eq. \eqref{eq:eta-w-eff}: they provide different ways to estimate the average powder catchment efficiency, respectively from dimensional and weight measurements. In particular, since $\bar \eta_h$ in bulk depositions is determined by the layer height $\bar h$, the study of powder catchment efficiency can be performed using a displacement sensor which measures the standoff distance $S$, defined as the distance between the deposition nozzle tip and the deposition surface. Accordingly $\bar h$ can be extracted differentiating $S$, as shown in the Supplementary Material.

	\subsection{Model for efficiency and height self-regulation}
	\label{sec:model3}
	Considering a LMD system based on a robotized deposition head, during a multi-layer deposition the robot height is incremented by $r_z$ for each layer, as represented in Figure \ref{fig:nozzle-track-scheme}(b-c). In an open-loop approach $r_z$ is kept constant. For an ideal process growth, such robot height increment should match the deposited layer height $\bar h$ to keep a constant standoff distance. However this condition is unlikely satisfied in real-world cases: a height mismatch $\Delta h = \bar h-r_z \neq 0$ is typically present if the process parameters are not carefully modeled or controlled. Moreover $\Delta h$ can change during the deposition due to the intrinsic process efficiency variability. A positive height mismatch $\Delta h>0$ corresponds to a growth which is faster than the robot movement along the build direction, and leads to a lower standoff distance; on the contrary, a negative mismatch $\Delta h<0$ leads to a greater standoff distance, hence to a departure of the deposition area from its initial relative position.
	
	A monotone mismatch $\Delta h \neq 0$ would cause a standoff distance divergence, eventually ending with a failure of the process itself when the deposition surface becomes too far or too close relatively to its optimal value. However, when the laser and powder beams are focused along the axial direction, variations in the standoff distance $S$ are also associated to modifications in the process conditions related to laser intensity and powder flux on the deposition area \cite{zhu_influence_2012}. In particular situations the presence of a non-zero gradient of $\Delta h$ with $S$ can lead to a stationary regime where $\Delta h=0$. This corresponds to a transient which brings the process growth to match the robot movements, hence to maintain a steady standoff distance \cite{haley_working_2019}. Process stabilization during the deposition of complex parts can be directly observed using inline optical monitoring \cite{donadello_monitoring_2019}. In general this mechanism is favorable, because it reflects passive process robustness against perturbations, and because it corresponds to enhanced geometrical accuracy since the actual deposition process tends to match the design model.
	
	The relation between deposition self-regulation and efficiency must be investigated in terms of energy and powder transverse distributions \cite{eisenbarth_spatial_2019}. The current work considers the case study configuration where the focal points of the laser and powder beams lay between the powder nozzle tip and the initial substrate position. In this common condition the open-loop self-stabilizing regime can be observed after a transient to smaller standoff distance values, i.e. with a positive layer mismatch $\bar h>r_z$. Such initially faster growth can be interpreted as the result of an increasing substrate temperature that favors powder melting at the beginning, in combination with a powder flux abundance which allows to deposit the first layers with high efficiency. Substrate temperature stabilization occurs typically after few layers \cite{huang_new_2019}, thus the effects of energy and powder spatial distributions dominate on the subsequent deposition growth rate, once a stable thermal condition is reached. The initial values of standoff distance and layer height increment also play an important role in the dynamics of process stabilization \cite{tan_process_2018}. It follows that the powder flux fraction available for catchment into the melt pool generated by the laser radiation changes while approaching the powder nozzle tip. This can introduce a powder catchment efficiency reduction, corresponding to a braking-like effect for the deposition growth, as a natural feedback mechanism which leads to a stationary regime where $\bar h=r_z$.
	
	Several approaches can be followed for modeling the deposition height in terms of laser-powder interaction  \cite{huang_comprehensive_2016,song_numerical_2018,kumar_determination_2013}. We propose a semi-empirical model for a quantitative interpretation of the process self-regulation, extending the results reported in other works \cite{thompson_overview_2015}. Accordingly the powder catchment efficiency will be derived with an explicit dependence on the standoff distance $S$ and on the process parameters, assuming a stationary deposition for bulk geometries.
	
	Here the theoretical powder catchment efficiency $\eta_{th}$ is defined as the product of two coefficients, one related to energetic factors $\eta_{en}$, the other depending on the powder-laser spatial interaction $\eta_{int}$:
	\begin{equation}
	\eta_{th} = \eta_{en}\eta_{int}\,.
	\end{equation}
	The laser-powder interaction is assumed to happen only within an area $A_{int}$, corresponding to the intersection between the respective beams on the deposition surface. This determines the effective powder mass rate that can be captured in the melt pool generated by the laser radiation. The approximation of uniform powder mass flux and laser intensity over the respective spot areas is introduced for simplicity. Therefore, if the powder flux is distributed over an area $A_p \leq A_{int}$, the spatial interaction coefficient is
	\begin{equation}
	\eta_{int} = \frac{A_{int}}{A_{p}}\,.
	\label{eq:eta-interaction}
	\end{equation}
	
	From an energetic point of view, efficiency is typically observed to grow while increasing the laser power, until a saturation condition is reached, with an effect on the substrate dilution phenomenon \cite{kaplan_process_2001,zhong_experimental_2015}. Accordingly, here $\eta_{en}$ is modeled as the fraction of powder flux which can be melt by the energy provided over the interaction area $A_{int}$. Specifically $\eta_{en}$ corresponds to the ratio between the laser power $P$ and the critical power $P_0$, the latter representing the minimum energy rate that would be required to fuse the incoming powder flow rate. Since both $P$ and $P_0$ should be normalized to $A_{int}$, $\eta_{en}$ does not depend on the spatial interaction, which characterizes the term $\eta_{int}$. A conservative estimation of the critical power $P_0$ can be obtained neglecting the energy required for the substrate melting during the fusion bond with the powder particles: this contribute is small in the assumption that the substrate is in a stationary thermal condition after being preheated by the previous deposition layers, and that only a thin surface layer gets melted, hence with a reduced substrate dilution. Therefore:
	\begin{equation}
	P_0=\dot{m}_{tot} \frac{\mathcal{C}_p \Delta T + \mathcal{L}}{\mathcal{A}}\,,
	\label{eq:saturation-power}
	\end{equation}
	where $\mathcal{C}_p$ is the material specific heat, $\Delta T=T_{m}-T_0$ is the excursion between room temperature $T_0$ and melt pool temperature $T_m$, $\mathcal{L}$ is the fusion latent heat, and $\mathcal{A}$ is the material optical absorptance. In usual conditions $T_m$ can be roughly approximated with the material melting point. A gaussian sigmoid function is applied to $P/P_0$ to empirically express the energetic coefficient $\eta_{en}$:
	\begin{equation}
	\eta_{en} = 1-\exp\left(-\frac{P}{P_0}\right)\,.
	\end{equation}
	This allows for a simplified phenomenological description of the efficiency saturation that is typically observed for $P \rightarrow P_0$. Such analytical representation is convenient in view of numerical calculations, since it exhibits a smooth connection between $\eta_{en}$ in the linear scaling regime for $P \ll P_0$, while tending asymptotically to $1$ for $P\gg P_0$. Moreover, for $P \sim P_0$, the corresponding value of $\eta_{en}$ closely matches the efficiency calculated from the lumped heat capacity model in similar conditions \cite{steen_laser_2010}. It follows that the semi-empirical relation for the theoretical powder catchment efficiency becomes
	\begin{equation}
	\eta_{th} = \frac{A_{int}}{A_{p}} \left[ 1-\exp\left(-\frac{P\mathcal{A}}{\dot{m}_{tot} (\mathcal{C}_p \Delta T + \mathcal{L})}\right)\right]\,.
	\label{eq:eta-model}
	\end{equation}

	\begin{figure}[tb]
		\centering
		\includegraphics[width=0.55\linewidth]{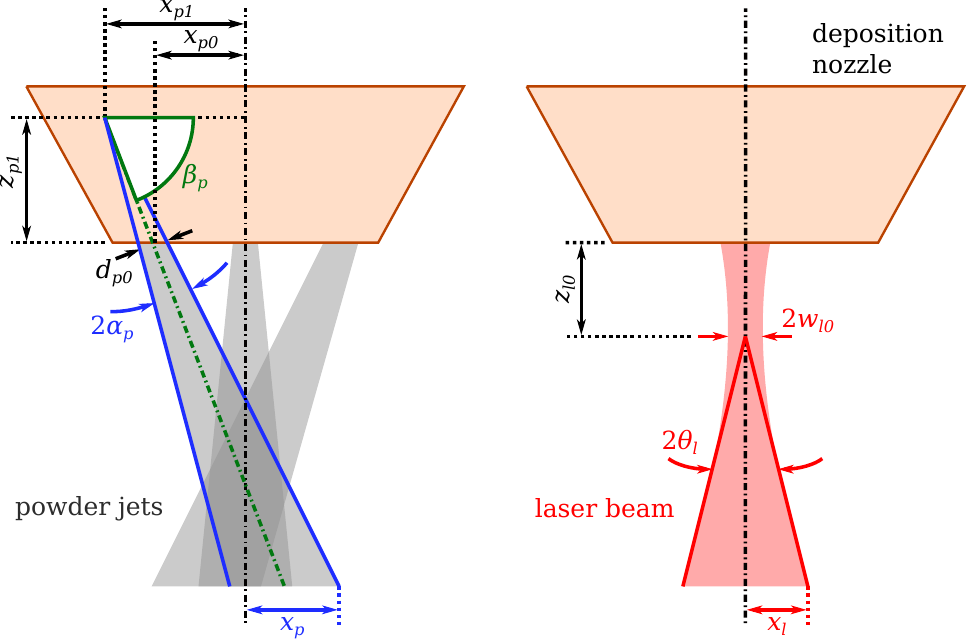}
		\caption{Side-view representation of the powder nozzle, with the main dimensions related to the $3$ powder jets and to the laser beam.}
		\label{fig:powder-geometry}
	\end{figure}

	An approximated geometrical model for the powder distribution can be provided for the calculation of $A_{int}$ and $A_{p}$, based on simple geometrical considerations for a specific deposition nozzle. Similar methods have been reported for the description of different configurations \cite{tan_development_2016}. Here a three-jet nozzle is considered, composed of $3$ powder streams arranged with axial symmetry at \ang{120}, as represented in Figure \ref{fig:powder-geometry}. The jets are tilted to form a hollow cone-like powder stream, coaxial to a focused laser beam passing along the nozzle axis. $A_{p}$ can be calculated as a function of the standoff distance $S$ knowing the divergence angle $2\alpha_p$ of the powder cones, the tilting angle of their axes $\beta_p$, and other geometrical parameters of the deposition nozzle. The details are reported in the Supplementary Material.

\begin{figure}[htb]
\centering
\includegraphics[width=1\linewidth]{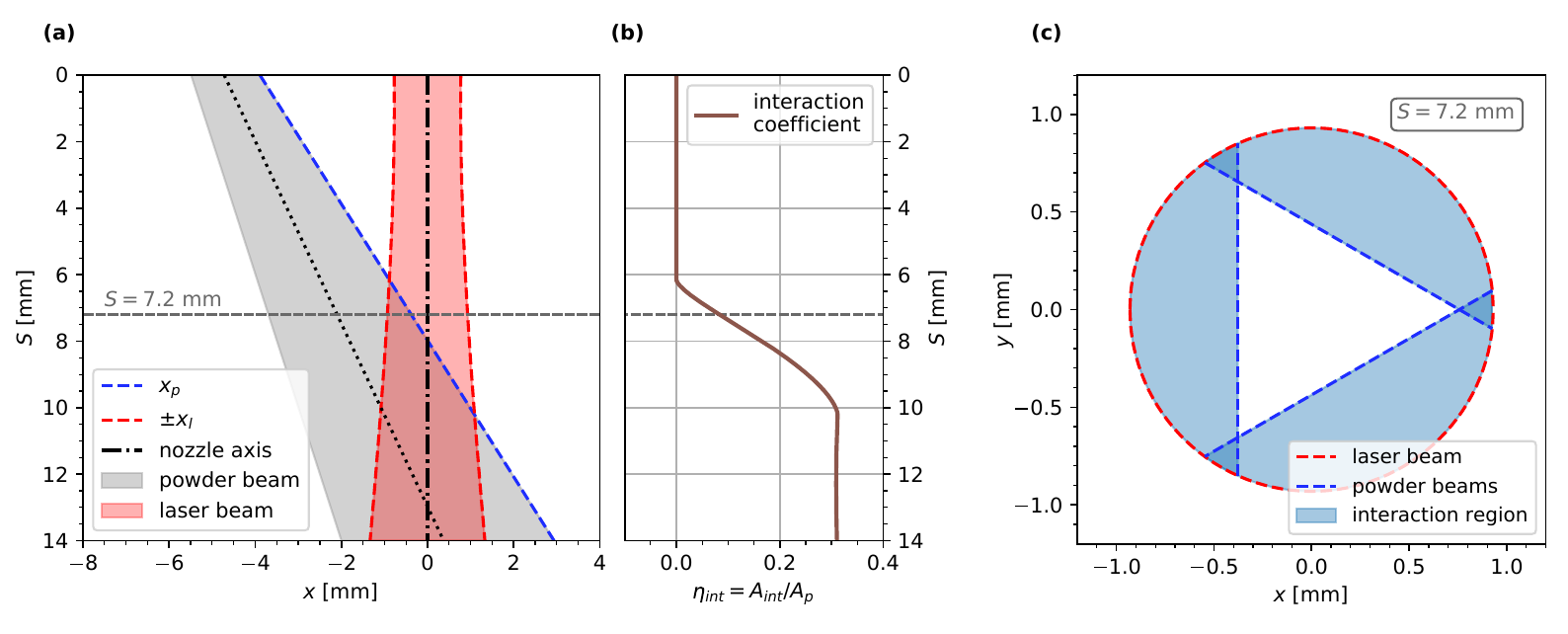}
\caption{(a) Quantitative representation of the coaxial laser beam and one of the powder beams; the origin is placed at the nozzle tip and the model correction factor is $c_g=2$. (b) Laser-powder interaction coefficient calculated as a function of standoff distance. (c) Transverse section of the intersections between laser and powder beams at an example standoff distance.}
\label{fig:powder-cone-vertical-section}
\end{figure}
	
	Figure \ref{fig:powder-cone-vertical-section} reports the quantitative calculations relative to $\eta_{int}$ for the nozzle geometry considered in the current study. The shaded regions in (c) represent the approximated laser-powder interaction area, taken at the example standoff distance represented by the dashed gray line in (a-b). The following regimes can be identified for the calculation of the interaction area $A_{int}$, depending on the standoff distance $S$.
	\begin{itemize}
		\item Closer than a minimum standoff distance value there is no intersection between the laser and powder spots, thus $A_{int}=0$ and the melt pool does not intercept the metallic powder flux.
		\item Above a characteristic standoff distance the laser spot fully overlaps with the powder area. If the powder spot diameter is bigger than the laser spot diameter the interaction area is limited by the laser beam itself. The initial standoff distance $S_0$ is typically chosen in this regime when precision depositions with good transverse resolution are required.
		\item In the intermediate region there is a strong dependence of $A_{int}$ on $S$, since only part of the melt pool area can intercept the powder flux. In a first approximation the interaction area can be calculated as the sum of $3$ circular segments. This assumption is valid if the laser spot diameter is significantly smaller than the elliptical axes of the powder cone sections. Such typical condition is verified in the considered configuration.
		\item Above a certain standoff distance another zone of partial intersection is present, tending to $A_{int}=0$. However this case is neglected in the current model since it lays far from the common working conditions, and because it corresponds to an unstable deposition regime.
	\end{itemize}
	
	The coordinates for the powder and laser spot borders relatively to the nozzle axis are represented in Figure \ref{fig:powder-cone-vertical-section}(a). These are defined as
	\begin{subequations}
		\begin{align}
		x_p &= \frac{S+z_{p1}}{\tan(\beta_p-\alpha_p)} - x_{p1}\\
		x_l &= c_g \sqrt{w_{l0}^2+\theta_l^2(S-z_{l0})^2}
		\end{align}
		\label{eq:powder-laser-coords}
	\end{subequations}
	in the plane of the respective powder and laser beam axes and referred to the nozzle tip center, with $x_{p1}$ and $z_{p1}$ depending on the nozzle geometry (see Supplementary Material). $x_p$ represents the inner cone border of the powder jet. The laser spot limit $x_l$ is assumed as the gaussian beam radius, and it can be calculated at the standoff distance $S$ knowing the $1/e^2$ beam waist radius $w_{l0}$, the beam half-divergence $\theta_l$, and the focus position $z_{l0}$ along the beam axis. A dimensionless geometrical factor $c_g$ is included in $x_l$: it summarizes the effects of the approximations introduced in the geometrical model, and in particular the discrepancy between laser spot and the actual melt pool size. This parameter will be determined \textit{a posteriori} by fitting the model predictions to the experimental results, finding values of about $2$. Such empirical calibration approach is convenient due to the complexity of providing a comprehensive quantitative model for the actual interaction between laser and powder beams. In fact it must be noted that $c_g$ can be influenced also by thermal accumulation effects, with a not trivial dependence on the chosen deposition strategy and parameters.
	
	The laser-powder interaction area is calculated from $x_p$ and $x_l$ considering the effect of superimposition for the $3$ powder jets:
	\begin{subequations}
		\begin{align}
		x_p \leq -x_l \; : & \quad A_{int}=0\\
		-x_l < x_p < x_l \; : & \quad A_{int} = 3\left(x_l^2 \arccos\left(-\frac{x_p}{x_l}\right) + x_p\sqrt{x_l^2-x_p^2} \right)\\
		x_p \geq x_l \; : & \quad A_{int}=3\pi x_l^2\,.
		\end{align}
		\label{eq:laser-powder-interact-area}
	\end{subequations}
	In the approximation of uniform powder and laser energy distributions within the respective spot areas, $\eta_{int}=A_{int}/A_{p}$ can be taken as the fraction of powder flow which can interact with laser beam, hence which can contribute to the effective melt pool and layer growth. This ratio depends on the standoff distance as plotted in Figure \ref{fig:powder-cone-vertical-section}(b) for $c_g=2$. Accordingly the theoretical powder catchment efficiency $\eta_{th}$ can be determined analytically as a function of $S$ for every process condition.
	
	From a different point of view, the theoretical layer height can be calculated from $\eta_{th}$ as a function of standoff distance exploiting Eq. \eqref{eq:eta-h-2}:
	\begin{equation}
	h_{th} = \eta_{th} \frac{\dot{m}_{tot}}{r_x v\rho}\,.
	\label{eq:h-th}
	\end{equation}
	This allows to simulate the standoff distance at layer $n$ as
	\begin{equation}
	S_{th,n} = S_0 - \sum_{i=1}^n(h_{th,i}-r_z)
	\label{eq:sod-simulation}
	\end{equation}
	considering the initial standoff distance $S_0$, and the $i$-th layer height $h_{th,i}$ being calculated at $S_{th,i-1}$. It can be observed that self-regulation can occur when the gradient $\partial h_{th}/\partial S>0$ and $h_{th,0}>r_z$, necessary conditions to match $h_{th}=r_z$. The proposed model represents an interesting tool for the prediction of the deposition growth for a given set of process parameters. These methods can allow for the analytical study of the self-stabilization mechanism, as well as for the efficiency optimization and the geometrical accuracy control.

	\section{Materials and methods}
	
	\subsection{LMD System}
	The LMD system is equipped with a fiber laser source (IPG Photonics YLS 3000), having \SI{1070}{nm} emission wavelength and \SI{3}{kW} maximum power. A \SI{400}{\micro\meter} fiber is employed to deliver the processing laser radiation to the deposition head (Kuka Industries MWO-I-Powder). The head is equipped with a \SI{129}{mm} focal length lens to collimate the laser beam, and with a \SI{200}{mm} focal length lens to focus it on the working area. The collimation is adjusted in order to have a \SI{1.2}{mm} laser spot diameter at the reference standoff distance $S_0=\SI{12}{mm}$. Optical monitoring devices can be integrated to the deposition head exploiting a dichroic mirror.
	
	The deposition head includes a three-jet nozzle (Fraunhofer ILT 3-JET-SO16-S) for powder delivery. The powder flow is controlled by a powder feeder (GTV TWIN PF 2/2‐MF). Argon is used as carrier gas to deliver the powder from the feeder to the nozzle. During the process argon is used also as shielding gas. The deposition head is mounted on a 6-axis anthropomorphic robot (ABB IRB 4600-45).

	\subsection{Powder and substrate materials}
	The employed powder is AISI 316L stainless steel, produced by Carpenter Additive. The powder chemical composition is reported in Table \ref{tab:powder-comp}. The physical properties of bulk AISI 316L are reported in Table \ref{tab:powder-prop}. The powder is characterized by a spherical shape, with a particle diameter in the range of $\num{45}-\SI{90}{\micro\meter}$. The powder was deposited on AISI 316L substrates having \SI{10}{\milli\meter} thickness and $\num{22}\times\SI{22}{\milli\meter\squared}$ dimensions, without any cooling system for the substrate.
	
	\begin{table*}
		\caption{Chemical composition of the employed AISI 316L powder.}
		\label{tab:powder-comp}
		\small
		\centering
		\begin{tabular}{l|lllllllll}
			\toprule 
			Element & Fe & C & Si & Mn & P & S & Cr & Mo & Ni \\
			Weight \si{\percent} & (base) & \num{0.023} & \num{0.36} & \num{1.30} & \num{0.016} & \num{0.005} & \num{16.96} & \num{2.45} & \num{10.89} \\ 
			\bottomrule 
		\end{tabular}
	\end{table*}
	
	\begin{table}
		\caption{Physical properties of solid AISI 316L stainless steel \cite{pinkerton_significance_2004}.}
		\label{tab:powder-prop}
		\centering
		\begin{tabular}{ll}
			\toprule
			Property & Value \\
			\midrule
			density $\rho$ & \SI{8e3}{\kilogram\per\meter\cubed} \\
			melting temperature $T_m$ & \SI{1670}{\kelvin} \\
			specific heat $\mathcal{C}_p$ & \SI{500}{\joule\per\kilogram\per\kelvin} \\
			fusion latent heat $\mathcal{L}$ & \SI{300}{\kilo\joule\per\kilogram} \\
			optical absorptance $\mathcal{A}$ & \num{0.35} \\
			\bottomrule 
		\end{tabular}
	\end{table}

	\subsection{Laser and powder beam characteristics}
	The laser beam parameters were characterized using a beam analyzer and are reported in Table \ref{tab:laser-beam-prop}. The focus position $z_{l0}$ is referred along the laser head axis, with the origin set at the powder nozzle exit.
	
	\begin{table}
		\caption{Characteristics of the processing laser beam.}
		\label{tab:laser-beam-prop}
		\centering
		\begin{tabular}{ll}
			\toprule
			Property & Value \\
			\midrule
			waist diameter $2 w_{l0}$            & \SI{0.77}{mm}          \\
			divergence angle $2 \theta_l$              & \SI{83}{\milli\radian} \\
			focus position $z_{l0}$            & \SI{0.9}{mm}           \\
			\bottomrule
		\end{tabular}
	\end{table}
	
	The powder beam parameters for the three-jet nozzle are reported in Table \ref{tab:powder-beam-prop}. The delivery channel inclination $\beta_p$, orifice diameter $d_{p0}$, and orifice radial position $x_{p0}$ were extracted from the nozzle geometrical model, defined relatively to the nozzle tip as represented in Figure \ref{fig:powder-geometry}. The powder dispersion $2\alpha_p$ was estimated by acquiring the powder beam stream with a high-speed camera (see Supplementary Material). In the considered configuration the position of laser beam waist and of the substrate in $S_0=\SI{12}{mm}$ lay above the convergence point of the powder streamlines.
	
	\begin{table}
		\caption{Characteristics of each jet contributing to the powder beam, referring to Figure \ref{fig:powder-geometry}.}
		\label{tab:powder-beam-prop}
		\centering
		\begin{tabular}{ll}
			\toprule
			Property & Value \\
			\midrule
			powder jet half-dispersion $\alpha_p$   & $\ang{6}$ \\
			delivery channel inclination $\beta_p$  & \ang{70}       \\
			delivery orifice diameter $d_{p0}$        & \SI{1.5}{mm}   \\
			delivery orifice radial position $x_{p0}$ & \SI{4.7}{mm}   \\ \bottomrule
		\end{tabular}
	\end{table}

	\subsection{System for real-time deposition mass measurement}
\begin{figure}[htb]
		\centering
		\includegraphics[width=0.55\linewidth]{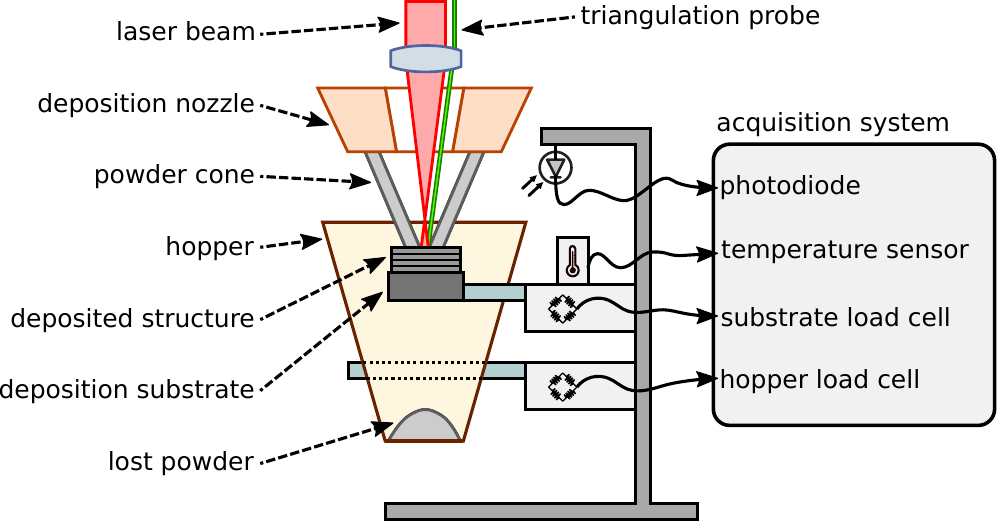}
		\caption{Scheme of the experimental setup for the real-time measurement of the deposited powder mass.}
		\label{fig:load-cell-setup}
	\end{figure}

	The system developed for the real-time measurement of the deposition mass is sketched in Figure \ref{fig:load-cell-setup}. The device is designed for independent measurements of the deposited and lost powder mass as a function of time during the deposition process. Such measurements are performed with two load cells based on strain gauges, that transduce the mechanical deformation induced by mass into a variation of electrical resistance. 
	
	The first load cell is directly connected to the deposition substrate to measure the powder mass which gets deposited. The second load cell sustains a hopper to measure the lost powder. The hopper is used also as a protection system to avoid accumulation of powder around the measurement device. The load cells are connected to the deposition substrate and to the powder hopper by insulating glass bars to suppress thermal conduction. Possible thermal effects which might perturb the acquisitions are monitored using an analog temperature sensor placed close to the load cells.
	
	The entire load cell system is protected with a plastic box, to avoid contamination and damaging of the electronic circuits. The box is purged by a continuous flow of argon, assuring gas recirculation and thermal stability. A photodiode is placed close to the deposition region to get a reference signal, for the synchronization of the mass measurement with the process laser emission. Moreover the measurement setup and electronics are protected from thermal irradiation using aluminum foils.
	
	The voltage signals from the load cells, the temperature sensor, and the photodiode are acquired at \SI{10}{\hertz} using an acquisition board (National Instruments USB-6009) connected to a computer running a LabVIEW program. The details related to the load cell system are reported in the Supplementary Material.

	\subsection{Deposition height monitoring with coaxial triangulation}
	The standoff distance was monitored in real-time during the deposition process using a custom coaxial triangulation setup. A detailed description of the measurement working principle was reported in a previous work \cite{donadello_monitoring_2019}. The optical scheme follows the former implementation, with some enhancements described in the Supplementary Material. The device is based on a low-power laser probe, superimposed to the deposition head optical chain exploiting a dichroic mirror. The probe beam reaches the \SI{200}{\milli\meter} lens of the deposition head in a slightly off-axis configuration, hence introducing a small beam deflection relatively to the optical axis. Then the beam passes through the nozzle towards the deposition region. A coaxial camera acquires the probe spot position on the melt pool, whose distance can be extracted using the triangulation principle.
	
	The probe spot position has been calibrated as a function of standoff distance \cite{donadello_coaxial_2018}, finding an enhanced sensitivity factor of \SI{0.106}{\milli\meter\per\pixel} over a \SI{20}{mm} range. The camera acquisition is performed at a frame rate of \SI{400}{\hertz}. The standoff distance is extracted in real-time using a Python algorithm. The same program is used to acquire the service variables of the robot system via an ethernet socket connection at a rate of about \SI{100}{Hz}. These are used to correlate the standoff distance data sets with the deposition head coordinates and with the laser emission.

	\section{Experimental campaign and sample analysis}
\begin{figure}[htb]
		\centering
		\includegraphics[width=0.35\linewidth]{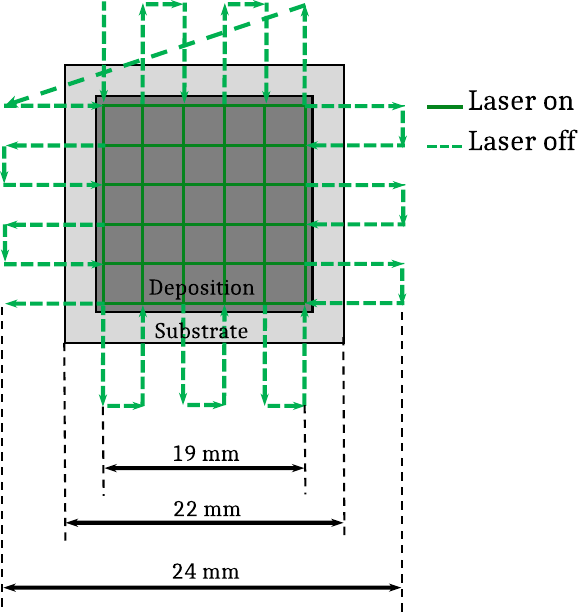}
		\caption{Deposition path strategy for building the square cuboids.}
		\label{fig:deposition-path}
	\end{figure}

	The main target of the experimental campaign was the investigation of the link between powder catchment efficiency and height growth in bulk geometries. Square cuboids having $\num{19}\times\num{19}\times\SI{10}{\milli\meter\squared}$ nominal dimensions were produced for the scope, obtained by means of multi-layer multi-track deposition. An alternated bi-directional deposition strategy was adopted, following the deposition path sketched in Figure \ref{fig:deposition-path}, with orthogonal scanning directions for adjacent layers.
	
	The experimental campaign followed a $2$-level factorial plan for $3$ varied parameters plus a central point, corresponding to a total of $9$ combinations. The varied parameters were the laser power $P$, the scanning speed $v$, and the delivered powder flow rate $\dot{m}_{tot}$. The central point was chosen as a typical working condition, optimized for bulk depositions with good dimensional resolution, yet representing a compromise for a reduced powder catchment efficiency. A number of $3$ replicates was performed for each deposition condition, for a total of $27$ randomized specimens. The fixed and varied parameters of the campaign are reported in Table \ref{tab:process-parameters}. Each combination of parameters is identified by a progressive label as reported in Table~\ref{tab:condition-labels}.
	
	\begin{table}
		\caption{Varied and fixed parameters of the experimental campaign.}
		\label{tab:process-parameters}
		\small
		\centering
		\begin{tabular}{ll}
			\toprule
			Parameter               & Value  \\
			\midrule
			laser power $P$                 & \num{525}, \num{612}, \SI{700}{\watt}                   \\
			scanning speed $v$              & \num{22}, \num{32}, \SI{42}{\milli\meter\per\second} \\
			powder feed rate $\dot{m}_{tot}$ & \num{0.11}, \num{0.16}, \SI{0.22}{\gram\per\second}        \\ 
			
			robot transverse increment $r_x$            & \SI{0.45}{mm}                \\
			robot height increment $r_z$              & \SI{0.2}{mm}                 \\
			total layer number & $50$ \\
			track number (for each layer) & $42$ \\
			initial standoff distance $S_0$ & \SI{12.0}{mm}                \\
			laser spot diameter (at $S_0$)    & \SI{1.2}{mm}                 \\
			carrier gas flow rate             & \SI{7.5}{\liter\per\minute}  \\
			shielding gas flow rate           & \SI{25.0}{\liter\per\minute} \\ \bottomrule
		\end{tabular} 
	\end{table}
	
	\begin{table}
		\caption{Labels of the considered process conditions.}
		\label{tab:condition-labels}
		\small
		\centering
		\begin{tabular}{llll}
			\hline
			Label & $P$ [\si{\watt}] & $\dot{m}_{tot}$ [\si{\gram\per\second}] & $v$ [\si{\milli\meter\per\second}] \\ 
			\hline
			{\color[HTML]{ff7f0e}$\blacksquare$} C1 & 612 & 0.16 & 32 \\
			{\color[HTML]{2ca02c}$\blacksquare$} C2 & 700 & 0.22 & 22 \\ 
			{\color[HTML]{d62728}$\blacksquare$} C3 & 700 & 0.22 & 42 \\ 
			{\color[HTML]{9467bd}$\blacksquare$} C4 & 700 & 0.11 & 22 \\ 
			{\color[HTML]{8c564b}$\blacksquare$} C5 & 700 & 0.11 & 42 \\ 
			{\color[HTML]{e377c2}$\blacksquare$} C6 & 525 & 0.22 & 22 \\ 
			{\color[HTML]{7f7f7f}$\blacksquare$} C7 & 525 & 0.22 & 42 \\ 
			{\color[HTML]{bcbd22}$\blacksquare$} C8 & 525 & 0.11 & 22 \\ 
			{\color[HTML]{17becf}$\blacksquare$} C9 & 525 & 0.11 & 42 \\ 
			\hline
		\end{tabular}
	\end{table}
	
	During the deposition the powder flow was kept continuous due to the slow response dynamics of the powder feeder. On the contrary, the laser operation was intermittent, with the emission enabled only for the intervals over the deposition substrate. The intervals with the laser turned off were necessary to avoid transitory effects related to acceleration and deceleration of the robot at the substrate borders, even if they introduce an unavoidable inefficiency contribution. Specifically the laser emission remained inactive for about \SI{0.5}{s} between consecutive tracks, and for \SIrange{2}{6}{s} between consecutive layers. The robot transverse increment $r_x=\SI{0.45}{mm}$ was chosen to guarantee a sufficient overlap between consecutive tracks, considering the typical single track width of about $\SI{1}{mm}$ observed in the considered conditions. The height increment $r_z=\SI{0.2}{mm}$ represents an optimal value for precision depositions.
	
\begin{figure}[htb]
		\begin{center}
			\includegraphics[width=0.5\linewidth]{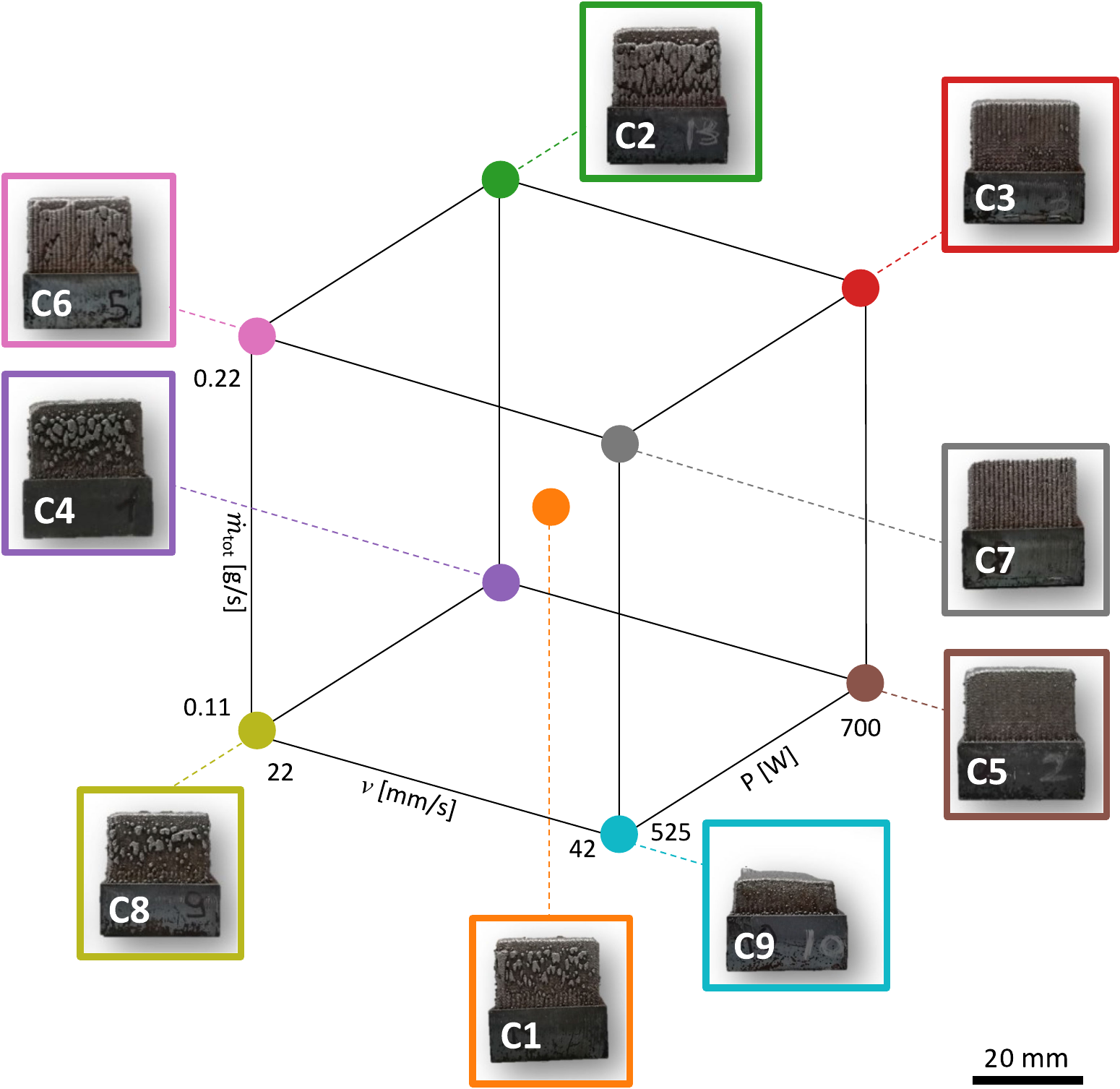}
		\end{center}
		\caption{Samples deposited in the different experimental conditions, represented in the space of the varied process parameters.}
		\label{fig:samples-cube}
	\end{figure}

	The experimental conditions of the campaign are represented in the space of process parameters in Figure \ref{fig:samples-cube}. Images of the realized samples are reported for each condition. From a qualitative point of view, the parts were deposited with a cubic shape as expected, except for condition C9: the latter represented an evident process failure, with an irregular shape and a significantly lower final height. Moreover it can be observed that the conditions at higher scanning speed showed more regular surfaces for the lateral walls, with a reduced debris attachment.
	
	\clearpage
	
	The deposition track shape was analyzed for few samples in significant process conditions, for a qualitative check of the model hypotheses and of the dimensional measurements. Accordingly the cuboids were cut along the growth direction to highlight the cladding structures. The cross sections were polished and chemical etched using a solution of distilled water, hydrochloric acid, and nitric acid for \SI{10}{\second}. The shapes of the deposition layers and tracks along the cross sections were analyzed using an optical microscope.

	\section{Results}
	
	\subsection{Mass measurements}
\begin{figure}[htb]
		\begin{center}
			\includegraphics[width=0.5\linewidth]{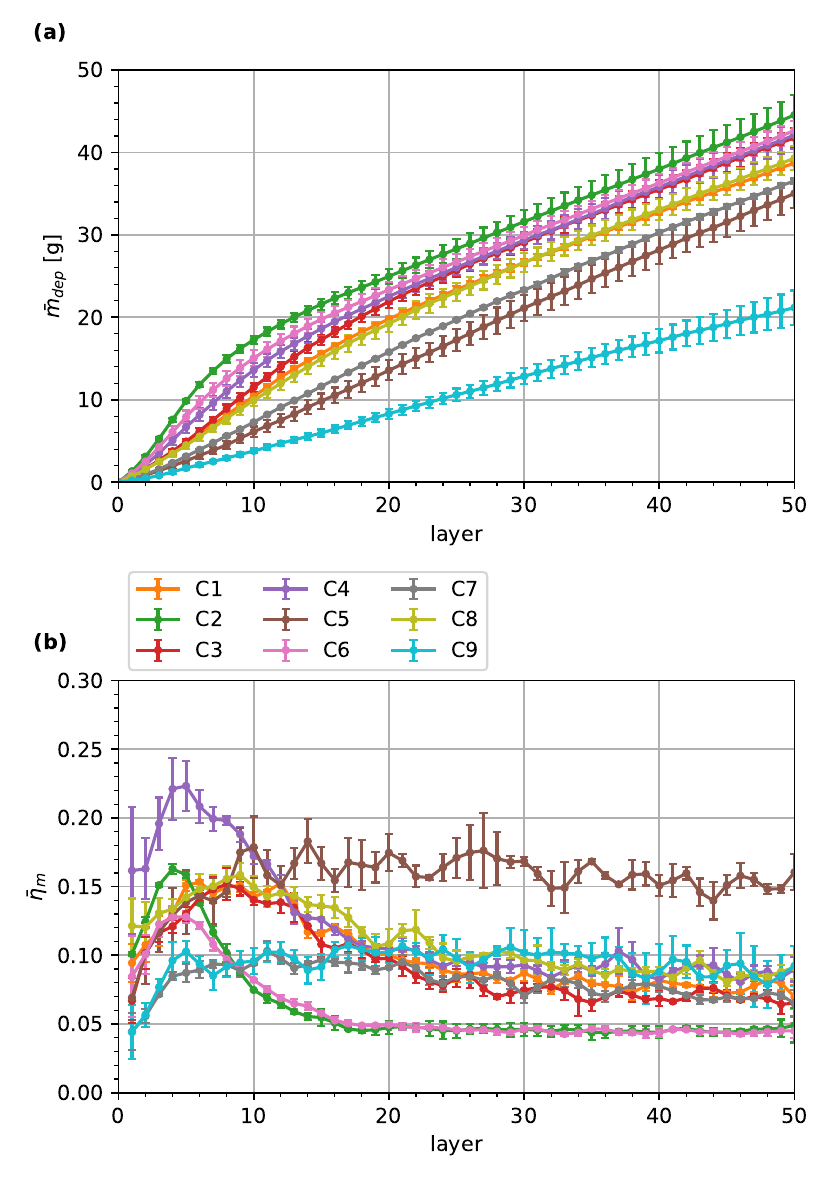}
		\end{center}
		\caption{(a) Average deposition mass, measured as a function of layer number and grouped by experimental condition. (b) Powder catchment efficiency, including the deposition path intrinsic inefficiency.}
		\label{fig:mass-eta-layers-grouped-conditions}
	\end{figure}

	The deposition mass was measured in real-time during the process by means of the load cell system. The initial mass offsets given by substrates and supports were subtracted from the load cell readouts. Each series was grouped by track and layer number, whose intervals were identified by acquiring the process laser emission with the monitor photodiode. The data have been smoothed using a $2$-nd order low-pass filter with $5$ layer cutoff, to suppress intra-layer variability related to the robot positioning vibrations. Indeed the current study is focused on the overall behavior of the bulk deposition process, not on the single layer spurious fluctuations. Finally, the data of each group of $3$ replicates have been averaged point-by-point to extract the mean trend for each experimental condition. The results for the average deposition mass $\bar m_{dep}$ are reported in Figure \ref{fig:mass-eta-layers-grouped-conditions}(a) as a function of layer number. Error bars represent the standard deviation within the replicates for each process condition.
	
	Conversely to the substrate load cell, the hopper sensor readout showed big oscillations, probably caused by vibrations induced by the robot movements. Accordingly the acquisitions of the lost powder were used only to check its final mass value, confirming that inefficiency is mainly determined by the powder that is not captured by the melt pool and falls from the substrate, being collected by the hopper. The powder catchment efficiency of each layer was calculated using Eq. \eqref{eq:eta-w-1} by differentiating $\bar m_{dep}$, assuming the total mass flow rate $\dot{m}_{tot}$ as a constant parameter, known from a preliminary offline calibration of the mass flow rate provided by powder feeder. The corresponding powder catchment efficiency curves are reported in Figure \ref{fig:mass-eta-layers-grouped-conditions}(b).
	
	The experimental results show that, in most conditions, the trend  of $\bar m_{dep}$ undergoes a slope change within the first $\sim10$ layers. After that, the deposition mass curves grow almost linearly with similar slopes, apart from condition C9 which departs with a clearly slower deposition rate. This behavior reflects on efficiency, with $\bar\eta_{m}$ reaching its maximum value $\lesssim 0.25$ within the initial layers for almost every condition. After such transient $\bar\eta_{m}$ stabilizes at lower values between $0.05$ and $0.15$, with an essentially constant trend for all the conditions.
	
\begin{figure}[htb]
		\begin{center}
			\includegraphics[width=0.85\linewidth]{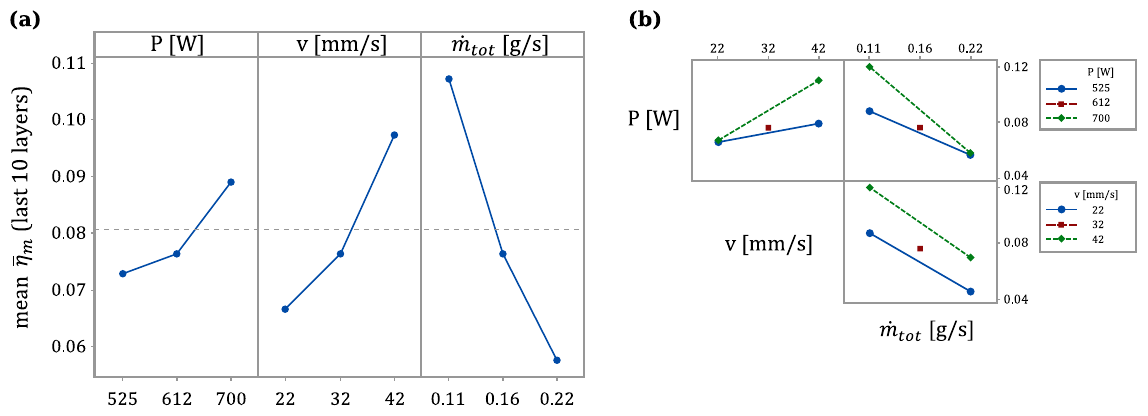}
		\end{center}
		\caption{Main effects plot (a) and interaction plots (b) for $\bar\eta_{m}$ averaged over the last $10$ layers.}
		\label{fig:main-effect-eta-prelim}
	\end{figure}

	A preliminary analysis is proposed considering the final values of the powder catchment efficiency, measured as $\bar\eta_{m}$ averaged over the last $10$ layers, to highlight the role of process parameters in a stationary regime. The main effects and interaction plots of the response variable are reported in Figure \ref{fig:main-effect-eta-prelim}. Each point in the plots represents the average for all the experimental runs taken at the corresponding parameter value, while the horizontal dashed line represents the average over all the experimental runs. The main effects plot implies that the strongest influence on $\bar\eta_{m}$ in the last layers derives from the delivered powder mass flow rate, $\dot{m}_{tot}$; nevertheless both laser power $P$ and transverse scanning speed $v$ affect the powder catchment efficiency in such stationary condition. While an increase in $P$ and $v$ reflects in a greater efficiency, higher values of $\dot{m}_{tot}$ appear to induce a lower $\bar\eta_{m}$. The interaction plots suggest interactions between processing laser power and deposition scanning speed ($P \cdot v$), and between processing laser power and delivered powder mass flow rate ($P \cdot \dot{m}_{tot}$). On the other hand, no strong interactions are expected between the scanning speed $v$ and the powder mass flow rate $\dot{m}_{tot}$. The respective analysis of variance (ANOVA) showed that all the parameters and their second order interactions are significant; the detailed results are omitted for the sake of brevity. 
	
	The analysis reported here can be interesting from a technological point view, since the final user is typically interested to the overall efficiency, measured as the deposit mass over the total deposition time. However it must be stressed that $\bar\eta_{m}$ is not relevant for a direct description of the deposition process physics, since it is comprehensive of losses related to the deposition strategy and, in particular, its value is reduced by the presence of intervals along the deposition path where the processing laser emission is disabled. Moreover the key role of standoff distance has not been taken into account at this preliminary stage. As it will be discussed in the following sections, the depositions in the different experimental conditions stabilize at different standoff distance values, hence in different conditions of laser-powder interaction.

	\subsection{Height measurements}
\begin{figure}[htb]
		\begin{center}
			\includegraphics[width=0.5\linewidth]{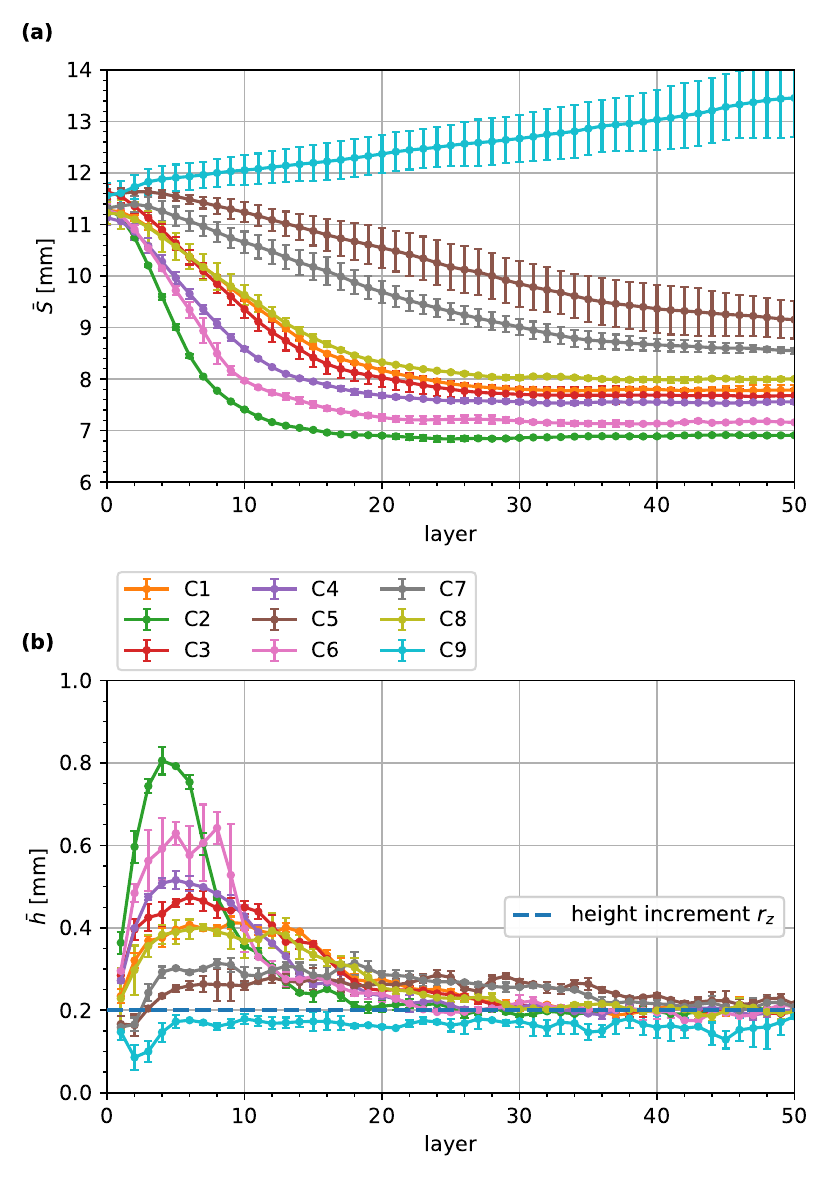}
		\end{center}
		\caption{(a) Average standoff distance, measured as a function of layer number and grouped by experimental condition. (b) Layer height calculated from standoff distance.}
		\label{fig:sod-height-layers-grouped-conditions}
	\end{figure}

	The standoff distance was measured in real-time as a function of deposition time by means of the coaxial triangulation system. The synchronous logging of the robot coordinates was used to identify the deposition head position, thus to extract the track and layer numbers. This allowed for a tomographic three-dimensional reconstruction of the deposited samples. Similarly to the analysis of $\bar m_{dep}$, the data sets of the mean standoff distance have been averaged over the $3$ replicates and low-pass filtered with a $5$ layer cut-off to suppress layer-to-layer fast fluctuations. The results for the mean standoff distance $\bar S$ in the different experimental conditions are reported in Figure \ref{fig:sod-height-layers-grouped-conditions}(a). The curves highlight that $\bar S$ departs from its initial value $S_0 \simeq \SI{12}{mm}$ with a slope which depends on the process condition. Apart from condition C9, all the curves show a reduction in the standoff distance, hence an actual deposit growth which is faster than the designed height. For almost every condition a stationary standoff distance is reached after the first $\sim 20$ layers, with the final values of $\bar S$ ranging between \SI{7}{mm} and \SI{9}{mm}, where the process enters in the self-stabilized regime for the deposition height growth.
	
	The average layer height $\bar h$ has been calculated differentiating the standoff distance (see the Supplementary Material). The results are reported in Figure \ref{fig:sod-height-layers-grouped-conditions}(b). For almost every condition the layer height undergoes an initial positive mismatch relatively to the robot increment along the build direction, fixed to $r_z=\SI{0.2}{mm}$ and represented as a horizontal dashed line in the plot. Such condition corresponds to a deposition height which grows faster than the programmed path, with $\bar h$ reaching values up to \SI{0.8}{mm} for condition C2, $4$ times $r_z$. After such transient the layer height tends to $r_z$, hence matching the robot height increment in the self-regulated regime. The only exception is C9, where a different behavior is observed, with the layer height being always slightly below $r_z$: this corresponds to process instability, with a departure towards bigger standoff distance values.
	
\begin{figure}[htb]
		\begin{center}
			\includegraphics[width=1\linewidth]{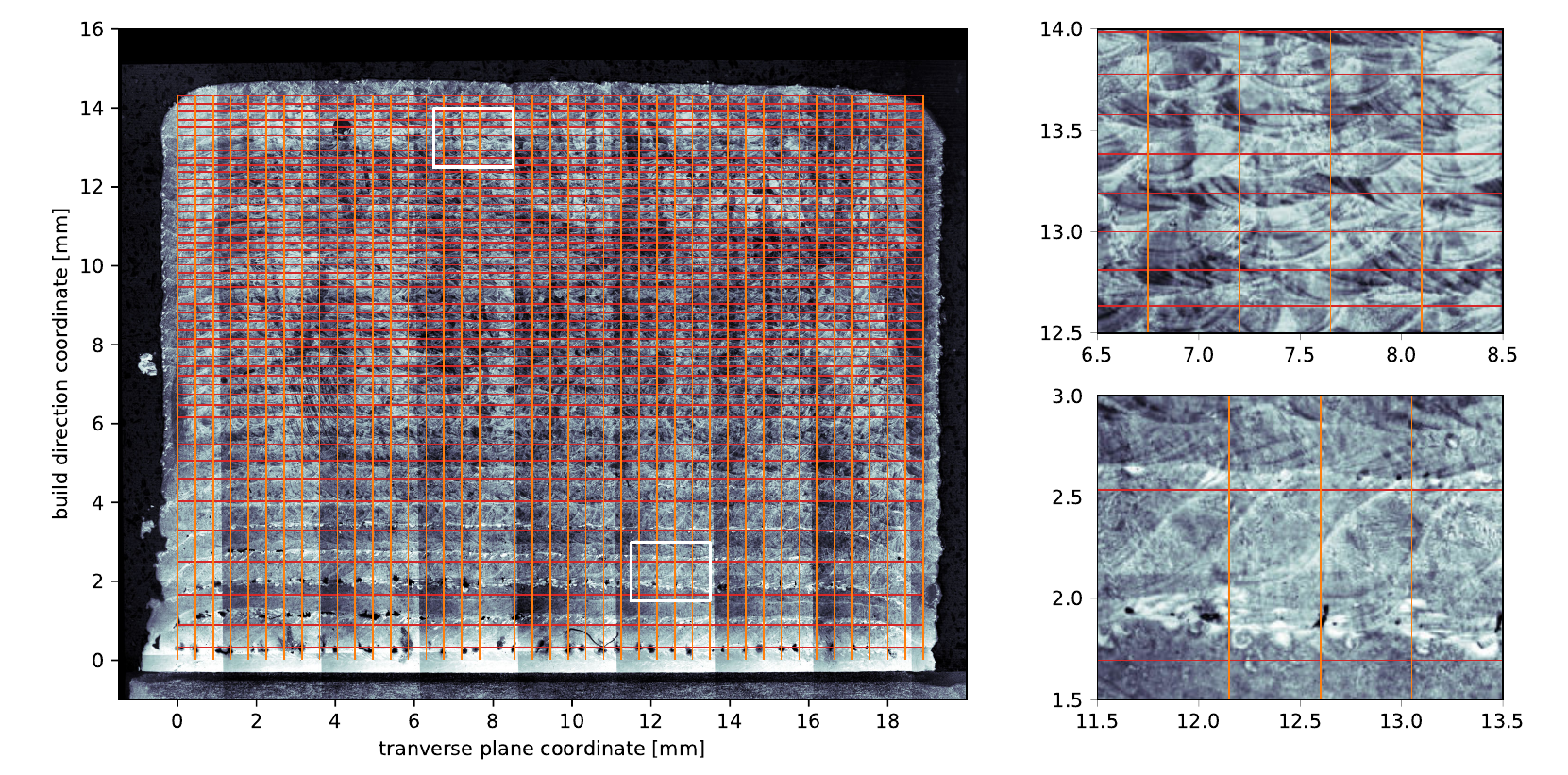}
		\end{center}
		\caption{Cross-section of a sample in the process condition C2. Horizontal lines represent the measured layer height, vertical lines represent the robot transverse increment. Some single tracks are magnified in the bottom.}
		\label{fig:section-layers-meas}
	\end{figure}

	Figure \ref{fig:section-layers-meas} reports the cross-section along the build direction for a sample deposited in condition C2. Horizontal lines are superimposed to the image to represent the average layer height measured with inline triangulation. Vertical lines having horizontal spacing equal to the robot transverse increment $r_x=\SI{0.45}{mm}$ are also reported. The picture provides a qualitative validation for the coaxial triangulation measurements for the description of the layer height during the deposition. Moreover, as it can be observed from the image magnifications, the deposition track shape changes significantly during the different process stages. Within the initial unstable layers the tracks are taller and strongly inclined, with the presence of pores caused by lack of fusion; conversely, the final self-regulated regime corresponds to a more regular track pattern, with the layer height matching the robot height increment $r_z$. These observations support the validity of the approximations which were introduced in Eq. \eqref{eq:area-cross-2} regarding the track cross-section geometry for a bulk deposition.
	
	\subsection{Powder catchment efficiency}
\begin{figure}[htb]
		\begin{center}
			\includegraphics[width=0.5\linewidth]{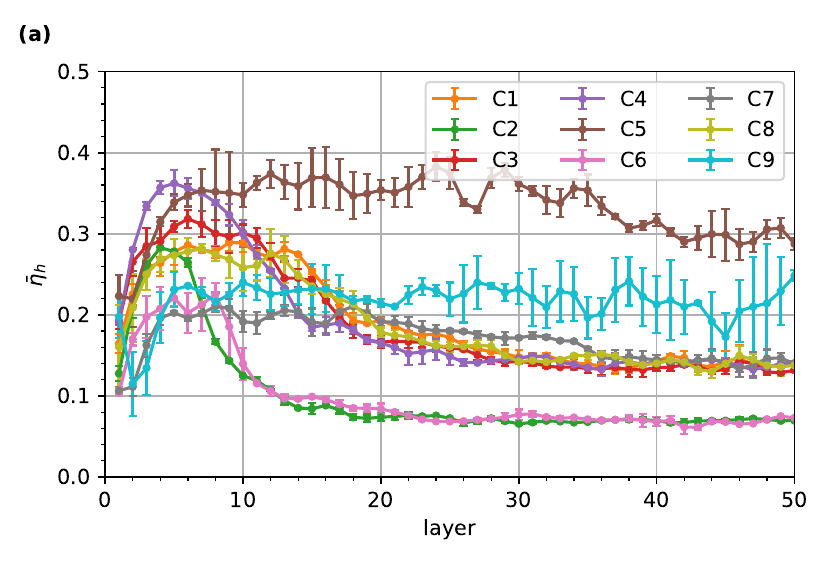}\\		
			\includegraphics[width=0.35\linewidth]{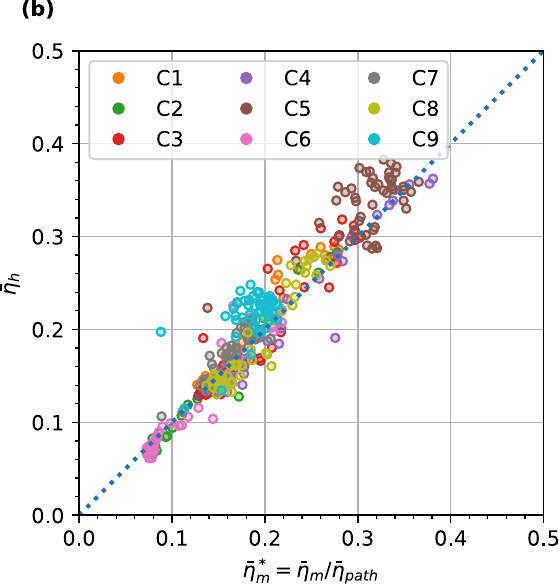}
		\end{center}
		\caption{(a) Geometrical definition of powder catchment efficiency, calculated from the layer height measurements. (b) Comparison between the effective powder catchment efficiency and its indirect geometrical measurement.}
		\label{fig:eta-mass-height-comparison-layer}
	\end{figure}

	The geometrical definition of powder catchment efficiency $\bar\eta_{h}$ can be calculated from the layer height measurements using Eq. \eqref{eq:eta-h-2}. The results are reported in Figure \ref{fig:eta-mass-height-comparison-layer}(a). They show that efficiency typically varies between $0.08$ and $0.4$, with an initial transient that follows the behavior which was already observed for $\bar\eta_{m}$ in Figure \ref{fig:mass-eta-layers-grouped-conditions}(b). The efficiency stabilizes after about $20$ layers, with C5 being the most efficient condition, while C6 and C2 correspond to the least efficient ones.
	
	For a quantitative comparison between $\bar\eta_{h}$ and the powder catchment efficiency $\bar\eta_{m}$ obtained from direct mass measurements\textbf{}, the latter must be corrected to isolate the effects related to the deposition process only. Indeed the definition of the overall powder catchment efficiency introduced in Eq. \eqref{eq:eta-w-1} includes an efficiency component $\bar\eta_{path}$ determined by the deposition geometry and by the robot dynamical parameters. Such path strategy coefficient has been estimated from the robot logs as the ratio between the laser emission intervals over the whole deposition time, following the definition of Eq. \eqref{eq:eta-path}. The values found for $\bar\eta_{path}$ are reported in the Supplementary Material, and showed a dependence on the scanning speed. The average value $\bar\eta_{path}=\num{0.54+-0.04}$ means that a remarkable powder fraction equal to $1-\bar\eta_{path}$ is lost during the robot position settlement, when the laser emission is turned off. 
	
	Although the inefficiency introduced by $\bar\eta_{path}$ is important from the point of view of the overall deposition costs, this is essentially determined by technical reasons, and it is not relevant for the comprehension of the deposition process physics. Therefore the effective powder catchment efficiency can be calculated as $\bar\eta_{m}^*=\bar\eta_{m}/\bar\eta_{path}$ as introduced in Eq. \eqref{eq:eta-w-eff}. The experimental values of $\bar\eta_{m}^*$ and $\bar\eta_{h}$ are compared in the scatter plot of Figure \ref{fig:eta-mass-height-comparison-layer}(b), where each point represents the average efficiency for a single layer. It can be seen that there is a good correspondence between the two efficiency measurements. Most of the experimental points lay close to the expected correspondence diagonal, with a Pearson correlation coefficient equal to $0.97$. This confirms that the powder catchment efficiency is linked to the deposition height as it was derived in the layer height model, demonstrating that $\bar\eta_{h}$ can be conveniently probed by means of an indirect geometrical measurement.
	
\begin{figure}[htb]
		\centering
		\includegraphics[width=0.85\linewidth]{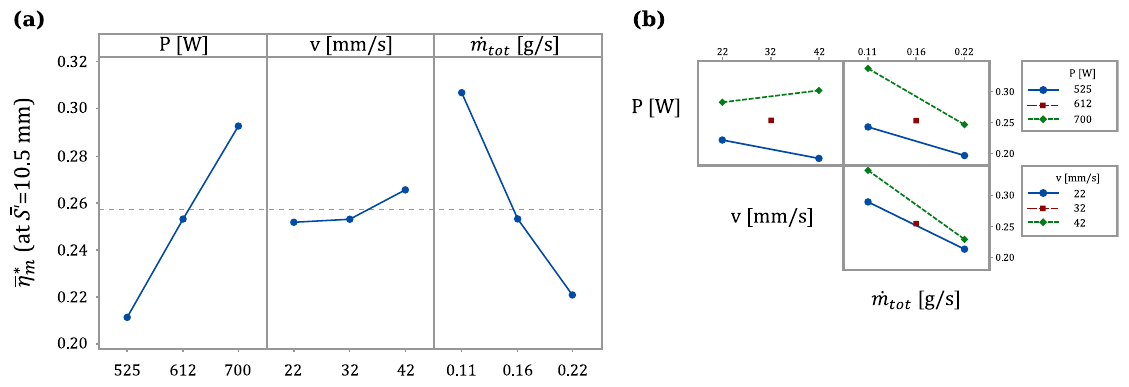}
		\caption{Main effects plot (a) and interaction plots (b) for the effective powder catchment efficiency interpolated at standoff distance $\bar S'=\SI{10.5}{mm}$. The unstable condition C9 was excluded from the analysis.}
		\label{fig:main-effect-eta}
	\end{figure}

	The results for the effective powder catchment efficiency $\bar\eta_{m}^*$ have been analyzed to identify the influence of the deposition parameters on the process physics. Differently from the results reported in Figure \ref{fig:main-effect-eta-prelim}, here the efficiency is analyzed at constant standoff distance $S$. This allows to suppress the strong dependence of the laser-powder interaction on $S$, which was discussed in the model for $\eta_{int}$. Accordingly $\bar\eta_{m}^*$ has been interpolated at the same standoff distance value for all the experimental curves: $\bar S'=\SI{10.5}{mm}$ has been conveniently chosen as test point, since such value was crossed for almost all the conditions after few initial layers and before self-stabilization. The substrate thermalization transients expected at the beginning are therefore avoided, and data comparison can be performed with consistency. Only the unstable condition C9 was excluded from the analysis, since it never crosses $\bar{S}'=\SI{10.5}{mm}$. The results for the interpolated $\bar\eta_{m}^*$ are reported in Figure \ref{fig:main-effect-eta} as a function of the varied parameters. The main effects plot shows that there is an influence of laser power and powder flow rate on efficiency: $\bar\eta_{m}^*$ grows with $P$, while it decreases with $\dot{m}_{tot}$. The ANOVA table, omitted for brevity, showed that only these two parameters are significant. Such results are in agreement with the predictions of the theoretical powder catchment efficiency $\eta_{th}$ introduced in Eq. \eqref{eq:eta-model}, essentially confirming the validity of the model hypotheses. Analogously, the lack of influence of $v$ on $\bar\eta_{m}^*$ is in accordance with the dependence of $\eta_{th}$ only on the ratio $P/\dot{m}_{tot}$, i.e. on the quantity of energy available for a given amount of delivered powder mass. Indeed the scanning speed is relevant for the initial transient caused by thermal accumulation and temperature rise at the beginning; however, in the stationary condition, $v$ affects equally the energy and powder concentrations, and its effect on the effective powder catchment efficiency can be neglected in a first-order approximation.
	
	\subsection{Effects of standoff distance on efficiency and height growth}
\begin{figure}[htbp]
		\centering
		\includegraphics[width=0.5\linewidth]{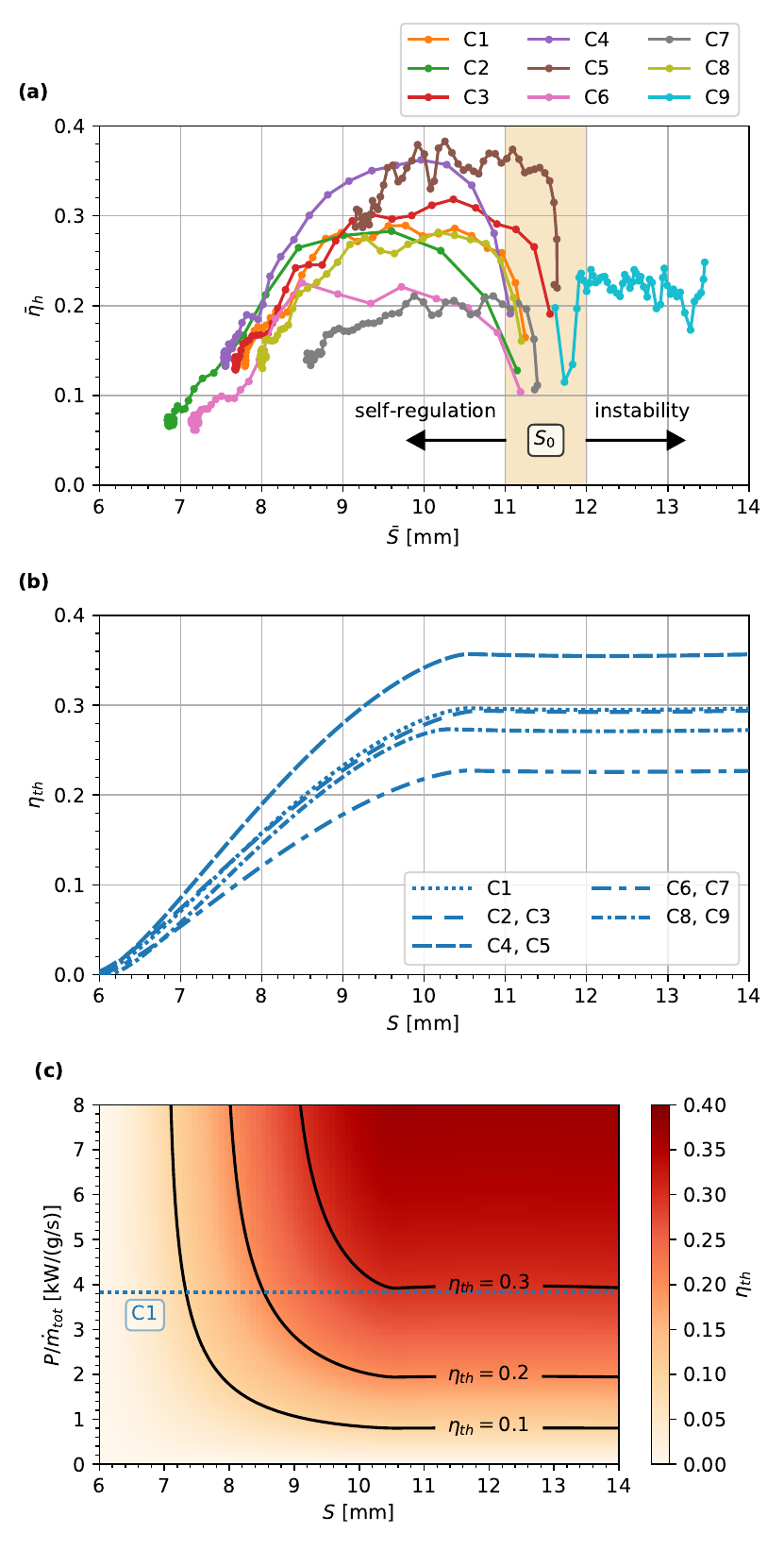}
		\caption{(a) Experimental powder catchment efficiency plotted as a function of standoff distance for the different experimental conditions. (b) Theoretical powder catchment efficiency calculated as a function of standoff distance. (c) Color map representing the theoretical powder catchment efficiency in the space of standoff distance and process parameters.}
		\label{fig:colormap_eta_th}
	\end{figure}

	The dependence of powder catchment efficiency on standoff distance is highlighted in Figure \ref{fig:colormap_eta_th}(a), where $\bar\eta_h$ is plotted as a function of $\bar S$. The deposition starts around the reference value $S_0\simeq \SI{12}{mm}$, in the shaded interval of the plot. The initial transient can be attributed to thermal stabilization effects. At the initial phase of the process starting at room temperature the substrate acts as a heat sink, dissipating the laser energy by conduction in a more marked manner. As the process proceeds over the layers, the substrate and the deposited material heat up, increasing the deposition efficiency until a stationary temperature is reached. Then the process tends to lower efficiency when it enters in the self-regulated regime, occurring at smaller standoff distance values and driven by the laser-powder spatial interaction. The only exception is condition C9, which departs toward bigger $\bar S$ values, hence to process instability. The highest efficiency in the self-regulated regime is observed for condition C5, which also corresponds to the smallest standoff distance mismatch relatively to its initial reference value $S_0$.
	
	In the proposed model the theoretical powder catchment efficiency $\eta_{th}$ was defined in Eq. \eqref{eq:eta-model} from the process parameters $P$ and $\dot{m}_{tot}$, with $\dot{m}_{tot}$ determining the critical melting power $P_0$ introduced in Eq. \eqref{eq:saturation-power}. Moreover $\eta_{th}$ depends on standoff distance $S$ as a consequence of the variable interaction area $A_{int}$, which can be calculated from Eq. \eqref{eq:laser-powder-interact-area} knowing the powder and laser beam parameters. Therefore the theoretical powder catchment efficiency $\eta_{th}$ can be calculated as a function of $\bar S$ for each process condition.
	
	It must be remarked that a geometrical correction factor $c_g$ was implicitly included in the definition of $\eta_{th}$ and introduced in Eq. \eqref{eq:powder-laser-coords}. This was necessary to take into account of the approximations assumed in the discussion, in particular regarding the laser-powder interaction region and spatial beam distributions. It can be expected that $c_g$ is influenced by the specific deposition geometry and strategy, determining the effects of thermal accumulation and melt pool evolution during the deposition. Accordingly $c_g$ was used as calibration parameter for the semi-empirical model, by fitting $\eta_{th}$ against the corresponding experimental efficiency curves $\bar\eta_{h}$ of Figure \ref{fig:colormap_eta_th}(a). The optimal values found for $c_g$ in the different conditions are reported in the Supplementary Material. The calibration factor exhibits a small dependence with the process parameters, and its average value over all the deposition conditions is $c_g=\num{2.19\pm 0.17}$. Since $c_g>1$, the model underestimates the effective laser-powder interaction region. Reasonably $c_g$ can be explained in terms of a melt pool which is bigger than the laser spot, e.g. considering an elongated melt pool, whose cue plays a relevant role in the powder catchment. It can be expected that $c_g$ depends on the process parameters that influence the melt pool temperature.
	
	The curves of $\eta_{th}$ are reported in Figure \ref{fig:colormap_eta_th}(b), calculated considering the respective $c_g$ values found from calibration. It can be observed that, although second-order corrections might be required for a quantitatively precise matching, the efficiency model is consistent with the measurements of Figure \ref{fig:colormap_eta_th}(a). Clearly the initial efficiency transients observed experimentally around $S_0$ are not reproducible, since the simulated efficiency curves assume a stationary thermal condition.
	
	From Eq. \eqref{eq:eta-model} it follows that the theoretical powder catchment efficiency is determined by two independent quantities, $S$ and the ratio $P/P_0$, which reduces to $P/\dot{m}_{tot}$ making explicit the dependence on the considered variable process parameters. Therefore $\eta_{th}$ can described in a bi-dimensional space as shown in Figure \ref{fig:colormap_eta_th}(c), where it is calculated considering the calibration coefficient $c_g=2.25$ corresponding to condition C1. The color map shows a high efficiency region for big values of $P/\dot{m}_{tot}$ around $S_0=\SI{12}{mm}$, where the laser intensity is sufficient to melt efficiently the available powder flux. Moreover efficiency decreases with $S$, due to the reduction of the laser-powder beam interaction. In such representation each process condition lays on a horizontal line, that is set by the values of $P$ and $\dot{m}_{tot}$, determining the efficiency evolution as a function of the variable standoff distance.
	
\begin{figure}[htbp]
		\centering
		\includegraphics[width=0.5\linewidth]{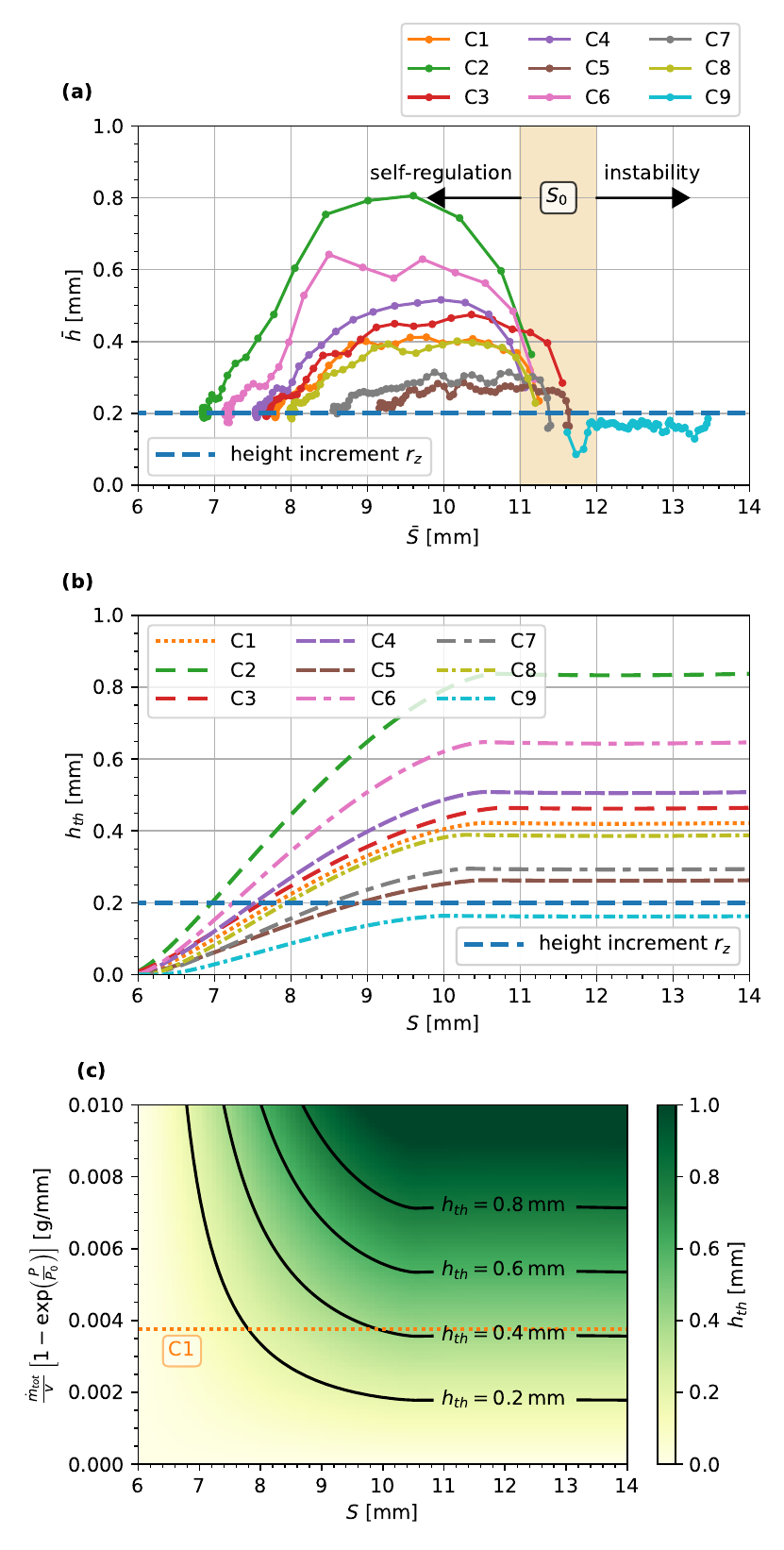}
		\caption{(a) Experimental layer height plotted as a function of standoff distance for the different experimental conditions. (b) Theoretical layer height calculated as a function of standoff distance. (c) Color map representing the theoretical layer height in the space of standoff distance and process parameters.}
		\label{fig:colormap_h_th}
	\end{figure}

	The self-regulating behavior is evident in Figure \ref{fig:colormap_h_th}(a), where the measured layer height $\bar h$ is plotted as a function of the average standoff distance $\bar S$. The robot height increment $r_z=\SI{0.2}{mm}$ is reported as a horizontal dashed line. It can be seen that, starting from the standoff distance $S_0\simeq \SI{12}{mm}$ within the shaded region of the plot, an initially positive height mismatch brings the deposition in the self-regulation regime, reached at lower $\bar S$ values. The corresponding height variability is influenced by the specific process condition. C2 shows the bigger variability and the smaller final standoff distance, while C5 is the condition which closely matches the robot height increment $r_z$, ending close to the initial standoff distance $S_0$. Remarkably, the reduced layer height variability observed for condition C5 is associated also to a high powder catchment efficiency, as commented for Figure \ref{fig:colormap_eta_th}(a). Only for condition C9 the process deviates to the unstable regime, with $\bar h<r_z$.
	
	The theoretical layer height is calculated using Eq. \eqref{eq:h-th} as a function of standoff distance in Figure \ref{fig:colormap_h_th}(b), considering the calibration coefficients $c_g$ that were previously determined. It can be seen that the results for $h_{th}$ are in good accordance with the experimental curves of Figure \ref{fig:colormap_h_th}(a), neglecting the transient around the initial standoff distance $S_0$ caused by thermal effects before substrate temperature stabilization. In Figure \ref{fig:colormap_h_th}(c) $h_{th}$ is calculated considering $c_g=2.25$, corresponding to condition C1. The theoretical height color map is represented as a function of two independent quantities, similarly to the treatment of $\eta_{th}$: the standoff distance $S$ (horizontal axis) and a combination of the parameters $\dot{m}_{tot}$, $v$ and $P$ (vertical axis), which can be obtained from Equations \eqref{eq:h-th} and \eqref{eq:eta-model}. Each parameter combination determines a process condition along a horizontal line. The process self-regulation is expected to happen along the isoline of the robot height increment, where the matching condition $h_{th}=r_z$ is fulfilled, with an evolution ruled by the non-zero gradient $\partial h_{th}/\partial S > 0$.
	
\begin{figure}[htb]
		\centering
		\includegraphics[width=0.5\linewidth]{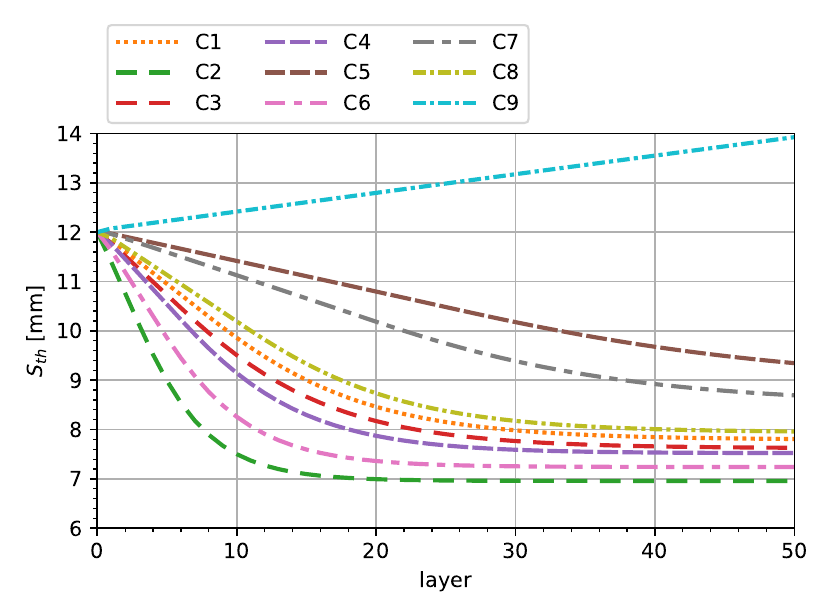}
		\caption{Evolution of the theoretical standoff distance simulated with the efficiency model.}
		\label{fig:sod-model}
	\end{figure}

	From a different point of view, the calculation of the theoretical efficiency $\eta_{th}$ allows to predict the deposition growth for each set of process parameters. This has been performed using Eq. \eqref{eq:sod-simulation} to simulate the standoff distance as a function of layer number, hence summing up the predicted layer height contributions $h_{th}$ starting from $S_0=\SI{12}{mm}$. This is equivalent to determine the process evolution along a horizontal line of the map reported in Figure \ref{fig:colormap_h_th}(c). The results corresponding to each considered experimental condition are reported in Figure \ref{fig:sod-model}. It can be seen that there is a good correspondence with the experimental standoff distance curves reported in Figure \ref{fig:sod-height-layers-grouped-conditions}(a). This means that the variable powder catchment efficiency, modeled in terms of process parameters and variable powder and energy distributions, allows to explain the experimental transient to closer standoff distances and lower efficiencies, hence providing a quantitatively valid interpretation for the process self-regulation and stabilization.
	
	\section{Discussion}
	The results for the deposition height and mass measurements showed that the powder catchment efficiency can be determined by measuring the layer height. In fact the $\eta_{m}^*$ and $\eta_{h}$ efficiency definitions resulted highly correlated, as evident from Figure \ref{fig:eta-mass-height-comparison-layer}(b) and predicted in Section \ref{sec:model2}. This confirmed the validity of the model hypotheses, stressing that these were introduced in the assumption of a stationary bulk deposition. The demonstration of $\eta_{h}$ as a good efficiency estimator allows for the possibility of using dimensional monitoring techniques also for the process optimization. Coaxial triangulation is particularly interesting for the scope, since its optical probe is essentially omnidirectional and can be easily implemented on a wide class of existing setups at a reduced cost: it allows flexibility in the deposition process monitoring, without requiring more complex and intrusive setups for the direct deposition mass measurement.
	
	The results highlighted that the deposition growth instability is mainly caused by efficiency variability. In fact, in stationary thermal conditions, the powder catchment efficiency depends on standoff distance due to the variable spatial interaction between the laser and powder beams, as shown in Figure \ref{fig:colormap_eta_th}(a). This allows to explain the self-regulation process as the balance between the initial heat accumulation and subsequent powder catchment efficiency reduction. In the considered configuration such passive stabilization mechanism is observed when the layer growth overcomes the robot height increment at the beginning, as reported in Figure \ref{fig:colormap_h_th}(a). However the standoff distance reduction reflects in a lower efficiency, acting as a negative feedback loop as predicted by the model and confirmed by experiments. This effect is ruled by the reduction of the laser-powder interaction below a characteristic standoff distance, as shown by the calculation reported in Figure \ref{fig:powder-cone-vertical-section}(b). At the end, the deposition stabilizes when the reduced efficiency brings the deposition layer height to match the robot height increment. This happens at different final standoff distance values, depending on the process parameters. 
	
	The deposition self-regulation is desirable because it corresponds to process robustness and geometrical accuracy enhancement. However it is in concurrency with powder catchment efficiency. A compromise between the two aspects must be taken into account. Moreover stabilization can be reached after a transient in the standoff distance, hence a departure from the geometrical design. Therefore the choice of the initial process parameters is particularly important for optimizing powder catchment efficiency, process stability, growth regularity, and part accuracy. The semi-empirical model for $\eta_{th}$ provides a powerful tool for this task. Such model was developed assuming a bulk deposition in a stationary thermal regime, characterized by multiple layers with a large number of tracks, and neglecting possible second-order effects determined by the specific deposition geometry or thermal drifts.
	
	After a preliminary model calibration through $c_g$, the theoretical efficiency map of Figure \ref{fig:colormap_eta_th}(c) allows to identify the most efficient parameter combination, related to the ratio between laser power and delivered powder rate, and the respective initial standoff distance. Analogously, the theoretical layer height map of Figure \ref{fig:colormap_h_th}(c) allows to identify the stability conditions depending on the required resolution, i.e. fixing the robot height increment and the scanning speed such that $h_{th}\ge r_z$. A wise choice takes the combination of initial standoff distance and robot height increment which lays close to the layer height predicted in the corresponding map region. This means that the process will stabilize soon, without varying significantly form the initial standoff distance and from the designed geometry, while maintaining an optimal efficiency. Moreover the isoline $h_{th}=r_z$ should be crossed transversely during the standoff distance evolution to reach stability, in a condition where the gradient $\partial h_{th}/\partial S>0$ is maximized to guarantee a robust and fast convergence.
	
	For the considered experimental campaign the best choice seems to be represented by condition C5, which is both efficient and characterized by a height growth that closely follows the programmed robot path. In fact the energetic coefficient $\eta_{en}$ is maximum for C5, since the high laser power and low powder mass flux guarantees a high melting rate, hence low powder losses. Furthermore the combination of these parameters with the respective scanning speed guarantees a layer height growth that almost matches the robot height increment at the initial standoff distance, hence process stability and deposition regularity. However it must be noted that, even in the most convenient standoff distance condition, the powder catchment efficiency is limited to less than $0.4$ by the partial interaction between the laser and powder beams, expressed by $\eta_{int}$ and calculated in Figure \ref{fig:powder-cone-vertical-section}(b). This is given by the laser spot size, which is chosen smaller than the powder beam spot to achieve a good transverse resolution.
	
	The model can be potentially used for an open-loop process optimization, since it predicts well the deposition growth as shown in Figure \ref{fig:sod-model}. Indeed the knowledge of the expected deposition growth could allow for a quantitative preliminary correction of the geometrical model to meet the design requirements. For example, this might be performed by compensating the deposition height mismatch in the programmed robot path, once the choice of the initial process parameters have been performed. Moreover the threshold between self-stabilized and unstable growth can be also identified from the model, reducing the risk of deposition failures and wastes. This represents a great advantage during the component design and realization, reducing the number of preliminary tests and manual operations required for the parameter optimization.

	\section{Conclusions}
	This work provides a systematic investigation of layer height variation and self-stabilization in LMD. Modeling and experimental methods have been developed to evaluate the powder catchment efficiency inline the process. The deposition mass and height have been monitored in real-time in different process conditions by means of a custom measurement system. The main outcomes of the work are as follows.
	\begin{itemize}
		\item The powder catchment efficiency mainly relies on the balance between laser power and delivered powder rate. The layer height depends on the interplay between powder catchment efficiency, delivered powder rate, and scanning speed.
		\item The standoff distance plays a crucial role in the determination of the powder catchment efficiency, hence of the layer height, because of the variable laser-powder interaction.
		\item The maximum achievable efficiency was limited by the partial overlap between laser and powder beams.
		\item The height growth deviates from the initial conditions, depending on the process parameters and on the amount of energy provided by the laser source.
		\item The self-stabilization mechanism is associated to efficiency reduction, which leads the deposition growth to match the height increment at different final values of standoff distance and powder catchment efficiency.
		\item Self-stabilization cannot be reached in conditions of low laser power, low powder flow rate, and high scanning speed, which do not guarantee a sufficient initial layer growth.
		\item The powder distribution, hence the nozzle design, plays a critical role on the selection of the optimal process parameters, especially of the initial standoff distance. 
		\item The use of a load cell system is a straightforward, but intrusive, approach to measure powder catchment efficiency. The coaxial triangulation system was proven as a convenient alternative tool for the process stability comprehension.
		\item Process parameter maps were calculated to establish the link between powder catchment efficiency, layer height, and standoff distance, and can be used to identify the optimal process conditions.
	\end{itemize}
	
	The presented methods might be applied to closed-loop control systems for obtaining a regular deposition growth. For example, the real-time dimensional measurement by means of coaxial triangulation, eventually combined with pyrometry for a convenient temperature measurement, can be integrated with a feedback correction on the process parameters, such as laser power, scanning speed, or powder flow. Nevertheless, some aspects remain open. The approach should be further applied to conditions of higher deposition rates. The developed analytical model is dependent on the nozzle type, which should be extended to coaxial nozzles as well. Finally, the inclusion of a melt pool geometrical model could overcome the requirement of calibration to the process parameters, providing a more general analytical description.
	
	\section*{Acknowledgments}
	The authors gratefully acknowledge Mr. Riccardo Caccia for his support during the experimental work. Mr. Eligio Grossi is acknowledged for his support in the design of the electronic acquisition setup, Mr. Stefano Mutti for his support in the realization of the communication software for the robot system. The authors would like to thank the BLM Group for the longstanding collaboration during the LMD cell development. The project presented in this paper has been funded with the contribution of the Autonomous Province of Trento, Italy, through the Regional Law 6/99 (Project LT 4.0). This work was supported by European Union, Repubblica Italiana, Regione Lombardia and FESR for the project MADE4LO under the call ``POR FESR 2014--2020 ASSE I -- AZIONE I.1.B.1.3''.
	
	\bibliography{bibliography}
	

\section*{Supplementary material}

# Supplementary Material for the article "Interplay between powder catchment efficiency and layer height in self-stabilized laser metal deposition"

Simone Donadello, Valentina Furlan,
Ali Gökhan Demir, Barbara Previtali

## 1 Details of the model

The main steps and quantities involved in the semi-analytical model described in the main text are summarized in Figure 1. Some specific aspects of the model are discussed in the following sections.

### 1.1 Deposition track cross-section

Referring to Figure 2, the cross-section area of a generic deposition track can be expressed as

$$A_{cs} = k_g w_t h_t \qquad (1)$$

where $h_t$ and $w_t$ are the single track height and width, while $k_g$ is a geometrical factor which depends on the track shape. E.g., in case of a half-elliptical track $k_g = \frac{\pi}{4}$, while for a rectangular track $k_g = 1$.

A requirement for a bulk deposition without empty volumes is that the single track width must exceed the path transverse increment $r_x$, thus $w_t > r_x$. This is a necessary condition for a partial overlap between each newly deposited track and the previous one within the same layer, introducing the overlapping factor as $0 < (w_t - r_x)/w_t < 1$. Therefore $r_x$ is relevant for the determination of the bulk cross-section area $A_{cs}$, since it influences the effective track width and height. This can be understood considering that, in the presence of track overlapping, the incoming powder flow contributes to the height growth over the whole deposition region, hence also over the previous track.

From purely considerations and in the assumption of a stationary process, i.e. with the track width $w_t$ and height $h_t$ not varying significantly within the same layer, the average contribution given by each track to the layer deposition is equivalent to a

1

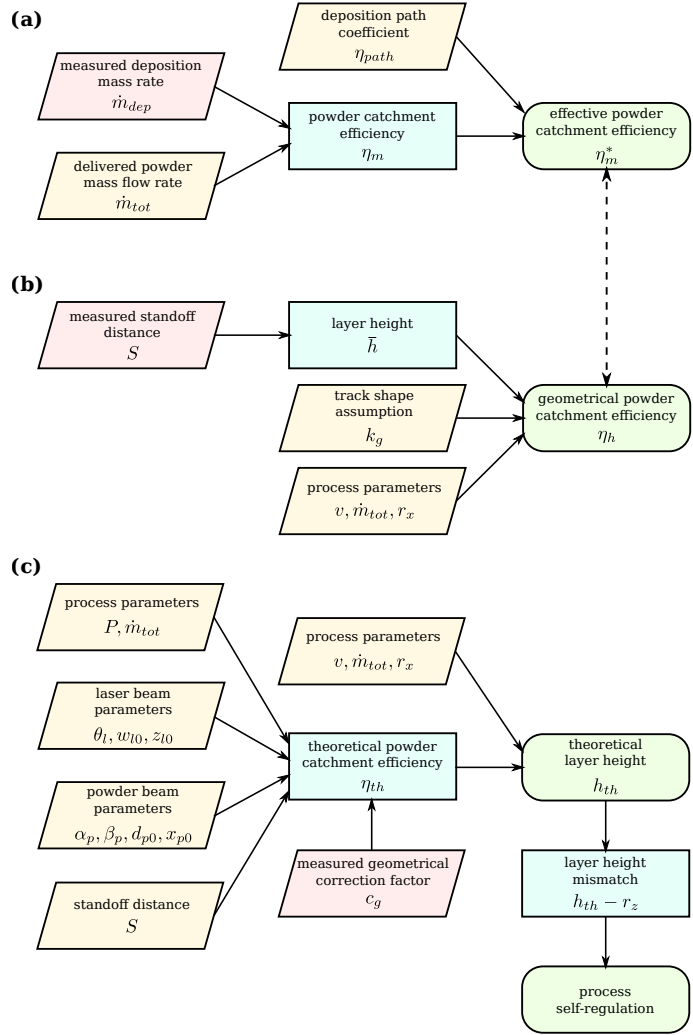

Figure 1: Main steps of the powder catchment efficiency models presented in the text.

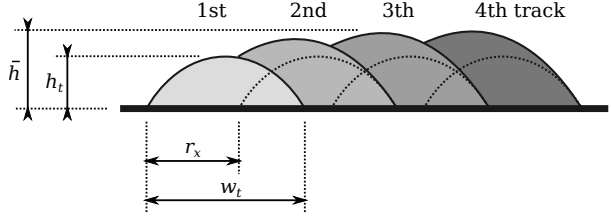

Figure 2: Qualitative cross-section representation of partially overlapping deposition tracks.

rectangular cross-section whose dimensions are set by the path transverse increment $r_x$ and by the average layer height $\bar{h}$. Therefore, under the previous hypotheses, $k_g = 1$ and the cross-section area of Eq. (1) averaged over a multi-track layer becomes

$$\bar{A}_{cs} \simeq r_x \bar{h} \,. \tag{2}$$

If $N_t$ is the track number for each layer, the error introduced by such approximation is of the order of $w_t \bar{h}/N_t$, whose contribution is negligible if the number of tracks which compose the layer is sufficiently big. This is a realistic condition for bulk depositions with $N_t \gg 1$.

## 1.2 Deposition layer height

Considering a multi-track multi-layer deposition, and referring to Figure 3, the average standoff distance $\bar{S}_n$ for the $n$-th layer can be expressed as

$$\bar{S}_n = S_0 + R_{z,n} - \bar{H}_n \tag{3}$$

where $S_0$ is the initial standoff distance value, $R_{z,n}$ is the relative robot height coordinate after $n$ layers, and $\bar{H}_n$ is the deposited structure height averaged within the considered layer. During the evolution of the deposition process the standoff distance can drift from its initial value due to instabilities in the deposition height, i.e. when the deposition path does not match the deposition growth and $R_{z,n} \neq \bar{H}_n$.

The deposit height $\bar{H}_n$ at layer $n$ is the sum of the layer height $\bar{h}_i$ for all the previous layers $i = 1 \ldots n$, thus

$$\bar{H}_n = \sum_{i=1}^{n} \bar{h}_i \,. \tag{4}$$

Analogously, the robot height coordinate $R_n$ at layer $n$ is the sum of the robot height increment $r_{z,i}$ for $i$ ranging from 1 to $n$. Here $r_{z,i} = r_z$ is considered a constant parameter during the whole deposition, thus

$$R_{z,n} = \sum_{i=1}^{n} r_{z,i} = n r_z \,. \tag{5}$$

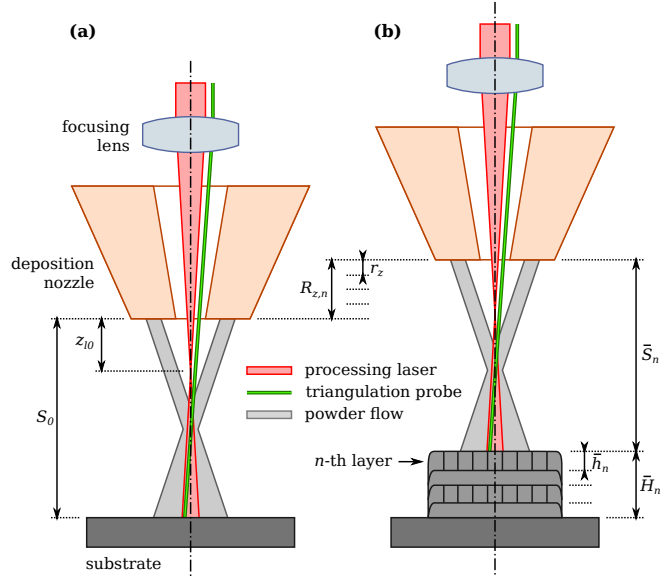

Figure 3: Schematic representation of the main dimensions related to a multi-layer deposition height in the initial condition (a) and after $n$ layers (b).

It follows that the $n$-th layer height can be calculated from standoff distance differentiation as
$$\bar{h}_n = \bar{H}_n - \bar{H}_{n-1} = r_z + \bar{S}_{n-1} - \bar{S}_n \,. \tag{6}$$

## 1.3 Powder beam geometry

The study presented in the main text considers a three-jet nozzle, composed of 3 powder streams arranged with axial symmetry at 120°. Each of the 3 powder jets can be seen as a cone having divergence angle $2\alpha_p$ outgoing from the nozzle tip as shown in Figure 4, with the cone axis tilted by $\beta_p$ relatively to the transverse plane. The diameter $d_{p0}$ of the powder delivery channel at the nozzle tip exit is assumed as a constrain for the powder cone basis at $S = 0$, whose center lays at distance $x_{p0}$ from the origin placed in the nozzle tip center. Since $\beta_p$, $d_{p0}$ and $x_{p0}$ are known from the nozzle geometry, while $\alpha_p$ can be determined experimentally by observing the powder flow with a camera, the coordinates of the hypothetical powder cone apex relatively to the nozzle tip center can be extrapolated as

$$\begin{aligned} x_{p1} &= \frac{d_{p0}}{2} \frac{\cos \beta_p}{\tan \alpha_p} + x_{p0} \\ z_{p1} &= \frac{d_{p0}}{2} \frac{\sin \beta_p}{\tan \alpha_p} \,. \end{aligned} \tag{7}$$

4

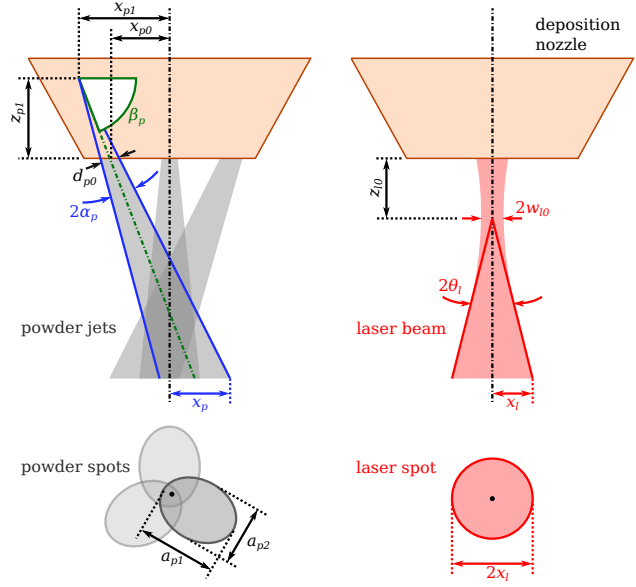

Figure 4: Qualitative side view of the powder nozzle, highlighting the dimensions related to the powder jets (left) and to the laser beam (right). The corresponding powder and laser spots in the transverse plane are represented in the bottom.

Under the previous assumptions and referring to Figure 4, it can be demonstrated that the major axis of the elliptical transverse section for each powder cone at a generic standoff distance $S > 0$ is

$$a_{p1} = (S + z_{p1}) \left( \frac{1}{\tan(\beta_p - \alpha_p)} - \frac{1}{\tan(\beta_p + \alpha_p)} \right). \tag{8}$$

The corresponding minor axis is

$$a_{p2} = 2(S + z_{p1}) \frac{\tan \alpha_p}{\sin \beta_0}. \tag{9}$$

Accordingly, the effective total area for the 3 elliptical powder sections is

$$A_p = \frac{3\pi}{4} a_{p1} a_{p2}. \tag{10}$$

5

# 2 Details of the experimental setup

## 2.1 Powder beam characterization

The dispersion angle of a single powder jet for the three-jet nozzle used in the LMD setup was determined experimentally by means of offline characterization, in absence of the processing laser emission. Indeed, the powder flow was acquired using a high-speed CMOS camera (Photron Fastcam Mini AX200) at 1 ms shutter time, considering the same conditions used for the experiments in terms of mass flow rate, carrier gas flow, and shielding gas flow. External illumination was used to highlight the trajectories of the powder particles, using a high-power white LED lamp. An example of image is reported in Figure 5. The half-dispersion $\alpha_p \sim 6°$ was estimated as the angle which contained the 98 % of the powder particles for a single jet.

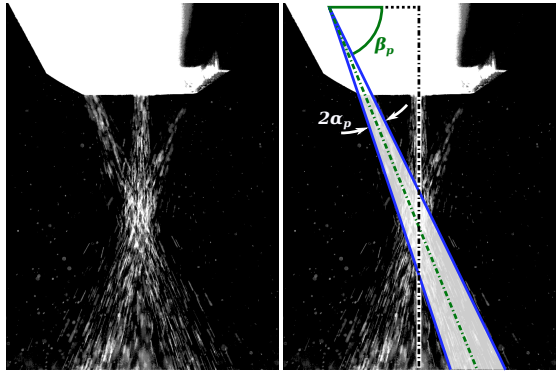

Figure 5: Image of the 3 powder jets spreading form the deposition nozzle. The characteristic angles for a single powder cone are highlighted on the right.

## 2.2 Mass measurement system

The system for the real-time measurement of the deposition mass is based on two load cell devices. Each load cell includes four strain gauges connected in a Wheatstone bridge configuration, insensitive to the distance of force application along the cell bar. The strain gauge bridge signal is amplified with an instrumentation amplifier circuit (Texas Instruments INA125P).

The dimension of the substrate is fitted to the programmed deposition path, allowing the unmelted powder to fall without accumulating on the substrate itself. The first load cell measures the mass of the substrate and of the deposited material, while the second load cell measures the mass of the lost powder collected by an hopper.

The hopper is made of brass sheet with 0.9 mm thickness. The choice of the material is related to the high heat dissipation of brass, which is important due to the high temperatures reached around the process region. The hopper implemented a self-balancing tip system, centered in the center of mass of a vessel which collected

the unmelted powder. The structure is also equipped with an anti-rotation system, consisting of an aluminum bar with two adjustable screws, that ensure greater stability during the deposition process by blocking excessive rotations of the brass vessel.

Thermal insulation has been treated carefully in the setup design. Indeed, the load cells can operate in a limited temperature range, up to 50 °C, and their readout can be strongly influenced by temperature variations. For this reason the load cells are placed far from the process region to avoid thermal damaging. The load cells are connected to the deposition substrate and to the powder hopper by insulating glass bars to suppress thermal conduction. Moreover the measurement setup is shielded from irradiation by aluminum foils.

The readout voltages of the load cells were calibrated using a set of standard masses, finding a proportionality factor of 0.086 V/g for the substrate cell, and of 0.0191 V/g for the hopper cell, on a scale ranging between 0 V and 10 V.

Possible thermal effects which might perturb the acquisitions were monitored using an analog temperature sensor placed close to the load cell. In fact a temperature dependence was observed above 30 °C while calibrating the load cell. However, thanks to the thermal shielding implemented in the setup, the temperature remained almost stable even in presence of the high-power processing laser radiation during long depositions. An average temperature of $(28.0 \pm 1.1)$ °C was maintained over the whole experimental campaign. This allowed to neglect the influence of ambient temperature on the mass measurements.

## 2.3 Coaxial triangulation system

The deposition height measurements were performed by means of triangulation system in a custom coaxial configuration. A 532 nm 4.5 mW probe laser is integrated in the deposition head optical chain, accessible through a lateral monitoring port. The collimated probe beam is deflected along the deposition optical axis by a dichroic mirror, transparent for the high-power laser wavelength, partially reflective in the visible and near-infrared (NIR) spectrum. The probe beam propagates to the 200 mm focal length lens of the deposition head in a slightly off-axis configuration, hence it gets focused with a non-zero deflection while exiting from the deposition nozzle orifice, as represented in Figure 3. This introduces a dependence of the transverse probe spot position on the standoff distance. The probe alignment offset of few mm relatively to the lens axis is chosen in order to get a small deflection angle, allowing to keep the probe spot within the melt pool considering the typical standoff distance variations during the deposition process.

The probe radiation which gets scattered from the melt pool surface is collected by the focusing lens, and back-deflected toward the monitoring port by the deposition head dichroic mirror. The beam crosses a further dichroic shortpass mirror with 650 nm cutoff, used for monitoring techniques in the NIR spectrum such as pyrometry or melt pool imaging, not considered in the current work. Half of the scattered probe radiation is deflected by a 50:50 beam splitter cube towards an iris used for spatial filtering. A 532 nm bandpass filter is used to eliminate spurious radiation. A compact Galilean-like telescope configuration is obtained by combining a 100 mm plano-convex lens with

a $-25\,\text{mm}$ bi-concave lens. Finally the probe spot is imaged on a 1.3 MP CMOS camera having $4.8\,\mu\text{m}$ pixel size (Ximea XiQ MQ013MG-ON).

The probe spot position has been calibrated as a function of known standoff distances, finding a sensitivity factor of $0.106\,\text{mm/pixel}$ over a 20 mm standoff distance range. The camera was triggered for the acquisition at 400 Hz frame rate, cropping the sensor to a $420 \times 96$ pixel area. The shutter time was limited for a short exposition time of 1 ms, thanks to the efficient light collection guaranteed by the $1''$ clear aperture of the monitoring optical chain.

The images were acquired by means of USB3 interface using a dedicated computer. A Python algorithm was used to extract the standoff distance in real-time during the process. The main steps performed for each frame are the following:

- the image is integrated along the short image axis, obtaining a 1D intensity vector along the probe beam deflection plane;
- the probe spot position and size are extracted efficiently as the vector distribution moments;
- finally the standoff distance is calculated applying a calibration curve, and stored for post-processing.

## 2.4 Positioning accuracy

An example of acquisition of standoff distance during the deposition of a cuboid sample is represented in 3D in Figure 6. It can be seen that the standoff distance departs from its initial nominal value $S_0 = 12\,\text{mm}$ after the first layers, with the color scale gradient showing an initially faster deposition growth and a final height mismatch of about 3 mm. The observed track-to-track and layer-to-layer variability can be ascribed to inaccuracies in the robot positioning and to laser head vibrations. These were quantified from the logs of the acquired robot axis encoders. The average standard deviation for the robot height within each layer was of the order of $70\,\mu\text{m}$; moreover, the robot height increment exhibited a standard deviation of about $30\,\mu\text{m}$ relatively to its nominal value $r_z = 0.2\,\text{mm}$, with peak mismatches up to 0.1 mm. Clearly these mechanical inaccuracies reflect on the optical distance measurement, influencing also the process conditions related to standoff distance. The robot positioning inaccuracies may play an important role at borderline conditions of process stability, since the standoff distance vibrations can eventually bring the deposition to operate in an actual condition of instability.

## 2.5 Model calibration and path efficiency

The parameters of the experimental conditions considered in the main text are reported in Table 1. The same table reports the calibration factor $c_g$, found by fitting the model for the powder catchment efficiency $\eta_{th}$ to the experimental data of $\bar{\eta}_h$ in the different conditions. Finally, the table reports the efficiency coefficient $\eta_{path}$ for $\eta_m$, related to the intervals of actual processing laser emission during the deposition time and measured from the robot control system.

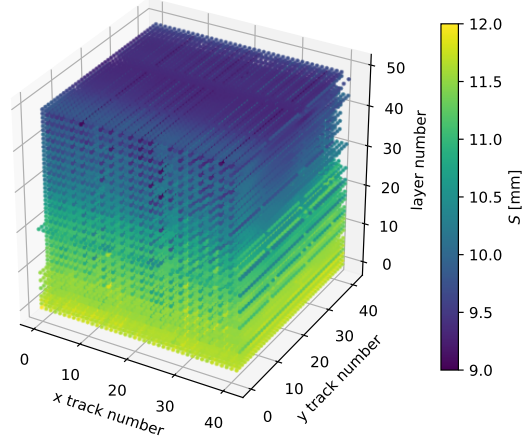

Figure 6: Three-dimensional representation of a deposited sample for the experimental condition C5 as defined in the main text. The color scale represents the measured standoff distance.

Table 1: Parameters of the considered process conditions, with the respective correction factors $c_g$ from model calibration and deposition path coefficients $\eta_{path}$ for the powder catchment efficiency.

| Label | $P$ [W] | $\dot{m}_{tot}$ [g/s] | $v$ [mm/s] | $c_g$ | $\eta_{path}$ |
|---|---|---|---|---|---|
| C1 | 612 | 0.16 | 32 | $2.25 \pm 0.02$ | 0.549 |
| C2 | 700 | 0.22 | 22 | $2.35 \pm 0.04$ | 0.586 |
| C3 | 700 | 0.22 | 42 | $2.42 \pm 0.02$ | 0.500 |
| C4 | 700 | 0.11 | 22 | $2.26 \pm 0.03$ | 0.586 |
| C5 | 700 | 0.11 | 42 | $2.24 \pm 0.02$ | 0.500 |
| C6 | 525 | 0.22 | 22 | $2.24 \pm 0.03$ | 0.586 |
| C7 | 525 | 0.22 | 42 | $2.09 \pm 0.02$ | 0.500 |
| C8 | 525 | 0.11 | 22 | $2.06 \pm 0.02$ | 0.586 |
| C9 | 525 | 0.11 | 42 | $1.85 \pm 0.01$ | 0.500 |


\end{document}